\documentclass{article}
\usepackage{PRIMEarxiv}
\usepackage{amsmath}
\usepackage[utf8]{inputenc} 
\usepackage[T1]{fontenc}    
\usepackage{hyperref}       
\usepackage{url}            
\usepackage{booktabs}       
\usepackage{amsfonts}       
\usepackage{nicefrac}       
\usepackage{microtype}      
\usepackage{lipsum}
\usepackage{fancyhdr}       
\usepackage{graphicx}       
\usepackage{xcolor}
\graphicspath{{media/}}     
\usepackage{multicol}

\title{A scaling model for measuring the morphology of African cities: Implications for future energy needs}

\author{
  Rafael Prieto Curiel\\
  Complexity Science Hub Vienna \\ Josefstaedter Strasse 39 \\ 1080 Vienna, Austria \\
  \texttt{prieto-curiel@csh.ac.at} \\
   \And
  Jorge E. Patino \\
  Research in Spatial Economics \\ Universidad EAFIT \\ Medellín, Antioquia, Colombia
  \And
  Brilé Anderson\\
  Sahel and West Africa Club - OECD \\ 2 Rue André Pascal \\ 75016 Paris, France
}

\begin{document}
\maketitle

\begin{abstract}

A large proportion of Africa's infrastructure is yet to be built. Where and how these new buildings are constructed matters since today's decisions will last for decades. The resulting morphology of cities has lasting implications for a city's energy needs. Estimating and projecting these needs has always been challenging in Africa due to the lack of data. Yet, given the sweeping urbanisation expected in Africa over the next three decades, this obstacle must be overcome to guide cities towards a trajectory of sustainability and resilience. Based on the location and surface of nearly 200 million buildings on the continent, we estimate the inter-building distance of almost six thousand cities. Buildings' footprint data enables the construction of urban form indicators to compare African cities' elongation, sprawl and emptiness. We establish the BASE model, where the mean distance between buildings is a functional relation to the number of Buildings and their average Area, as well as the Sprawl and the Elongation of its spatial arrangement. The mean distance between structures in cities -our proxy for its energy demands related to mobility- grows faster than the square root of its population, resulting from the combined impact of a sublinear growth in the number of buildings and a sublinear increase in building size and sprawl. We show that when a city doubles its population, it triples the energy demand related to commutes.

\end{abstract}

\section{Introduction} 

The world's future is in cities. Projections estimate almost 7 out of 10 people will live in a city by 2050. Whilst many parts of the world have already undergone urbanisation, the next three decades will bring sweeping changes in African cities. An additional 950 million will become urbanites by 2050 in Africa, compared to 574 million people in 2015. More people will need more buildings, future homes, schools, hospitals, markets, and all the other stops of daily life. Where and how these buildings are constructed matters since today's decisions will last for decades. The resulting morphology of cities -in other words, whether it is sprawling or compact, monocentric or polycentric, fragmented or contingent, densely or sparsely populated- has lasting implications for a city's energy needs \cite{bai2016defining, angel2020shape, seto2012global, parnell2018, oliveira2018worldwide, pumain2004scaling, li2020influence, zhou2017role, seto2014human}.

So far, efforts to tackle these questions have been waylaid by the lack of data. Isolated analyses of sprawl or fragmentation using population density data of major African cities exist, but smaller and intermediary cities are often excluded. This is troubling since most of the future urbanisation in the forthcoming decades will arise in these cities. Much of the collective knowledge of urban form in African cities and future energy needs are based on samples of only a few cities. However, recent advancements in data availability open new ways of observing patterns in cities at a global scale \cite{candipan2021residence, depersin2018global, BigDataPolicyProblems, Tusting2019, guan2020delineating}.

To address these data gaps, we capitalise on the newly available Google AI Africa Open Buildings dataset, which maps the location and area of every building on most regions of the continent (Appendix A). We combine the infrastructure data with building's height data modelled by the German Aerospace Centre (DLR) \cite{sirko2021continental, esch2022world}, with street network metrics \cite{OpenStreetMap} and with terrain metrics \cite{NASAJPL2020}. The granularity of these datasets, combined with prior work that maps the geographical location and boundaries of nearly 6,000 urban agglomerations \cite{Africapolis} and our models of urban form, it is possible to estimate African cities' transport requirements and energy needs with a never before seen accuracy.

\section{Results} 

\begin{figure*} \centering
\includegraphics[width=\textwidth]{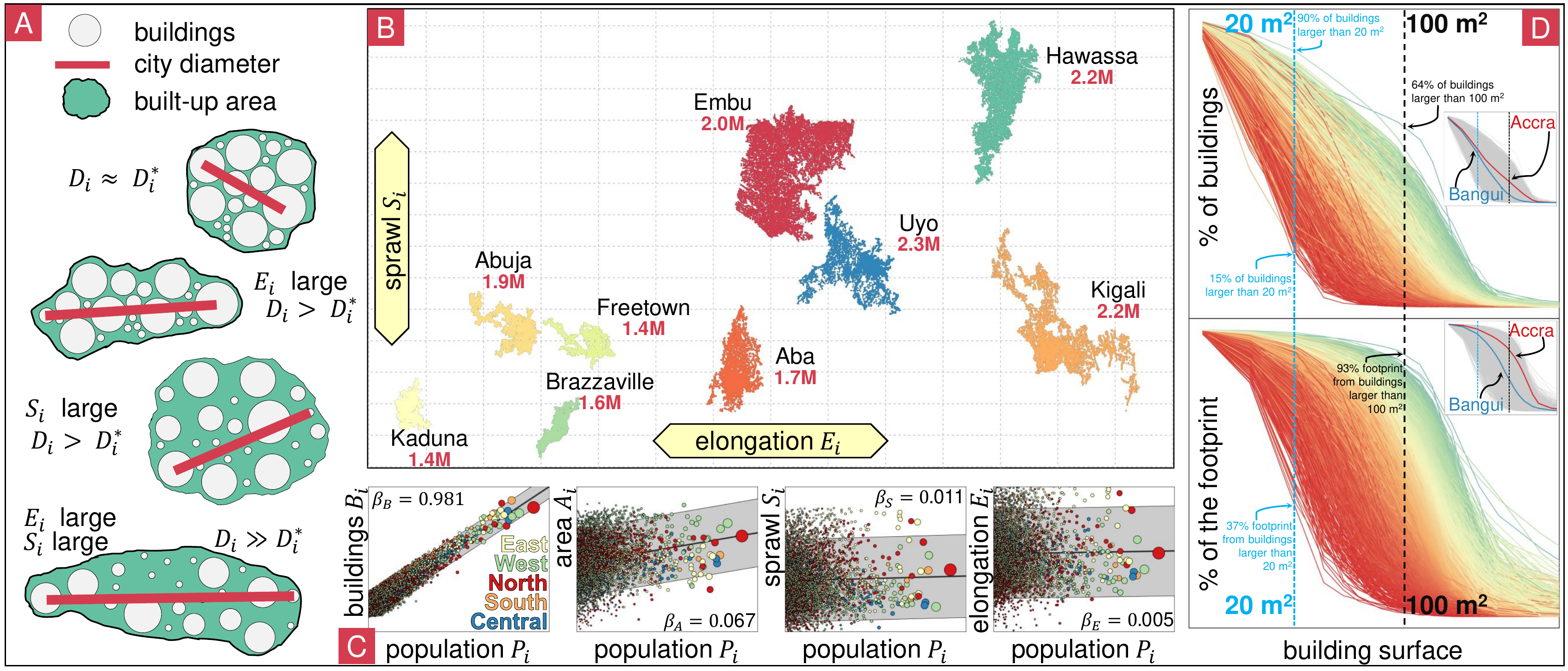}
\caption{A - Cities may be elongated, sprawling, or both, depending on the location of their buildings. B - Footprint of buildings observed across nine cities with a similar population (between 1.4 and 2.3 million inhabitants). Cities with an elongated shape have a larger diameter meaning that the mean distance between buildings is longer. Cities with more sprawl tend to have more space between buildings, contributing to longer distances. C - Observed increase in the number of buildings $B_i$, their area $A_i$, the sprawl $S_i$ and the elongation $E_i$ as the population of the city increases. D - Cumulative fraction of the number of buildings (top) and surface (bottom) of a city formed by buildings bigger than some area. Highlighted are the observed curves in Accra and in Bangui.}
\label{MsFigure1}
\end{figure*}

Here, we construct a set of indicators to characterise urban morphology based on the geographical distribution and size of millions of buildings in African cities \cite{ewing2014measuring}. The objective is to relate the morphology of cities to distance indicators (e.g., sprawl, elongation, polycentricity) and the consumption of commuting energy \cite{barthelemy2019statistical, zhou2017role, li2020influence}. We use the coordinates and surface of 183 million buildings constructed in nearly 6,000 cities in Africa (Appendix A). For each city identified by Africapolis \cite{Africapolis}, we measure the mean inter-building distance, which is our proxy for commuting energy consumption (Appendix A). Based on the expression of the average distance between two points inside a circle, a functional equation for the mean distance $D_i$ between two buildings inside a city $i$ is given by
\begin{equation} \label{MeanDist}
D_i = \frac{128}{45 \pi} \sqrt{ B_i A_i S_i E_i},
\end{equation}
where $B_i$ is the number of buildings, and $A_i$ is their average area. $S_i$ is the \emph{sprawl} index, which is the space between buildings, where smaller values come from a compact city. $E_i$ is the \emph{elongation} of a city (or anisometry \cite{zhou2017role}), where smaller values indicate a round shape and higher values suggest a more elongated footprint (Appendix B). The sprawl $S_i$ and elongation $E_i$ are comparable across cities of different sizes, based on the maximum distance between buildings (Figure \ref{MsFigure1}).

The BASE model decomposes intracity distances into four multiplicative components \cite{gudipudi2019efficient, angel2021anatomy}. The model gives an estimate for distances between buildings if a city is round and compact, $D^\star_i$, obtained in Equation \ref{MeanDist} with $E_i = S_i = 1$, reflecting how distances increase due to the city's footprint (number and size of buildings). We define the \emph{fragmentation} of a city as $\psi_i =D_i / D^\star_i =\sqrt{S_i E_i}\geq 1$, the ratio between the observed and minimum distances, in other words, the extent to which a city departs from a packed pattern (e.g., the space between buildings: green space, blue space, roads). More fragmented cities have longer distances due to factors other than more or bigger buildings. We use cities' elongation, sprawl, and fragmentation to detect what contributes to longer distances and characterise cities' morphology across Africa. The physical terrain and the political boundaries play a crucial role in how cities grow into elongated or sprawled patterns. Urban areas with greater altitude variation -an indicator of rugged terrain- tend to be more elongated and have greater sprawl than urban areas built on flat terrain (Appendix C). Rugged terrain and steep slopes pose a natural barrier to building roads or taller buildings along with other infrastructure. The urban footprint adjusts to the location of usable land, resulting in a more organic and less compact urban shape. Cities near an international border are growing at a faster speed than other cities, and they also tend to be more elongated \cite{BorderCitiesOECD}. More polycentric cities tend to have elongated shapes, greater sprawl and longer distances (Appendix D) \cite{RevealingCentrality, bartosiewicz2020investigating}.

The number of buildings within a city grows with population, with roughly one extra building every 2.6 people. However, that number decreases slightly with size, reflecting a shared infrastructure. At a continental level, a city with ten times the population has 9.6 times the number of buildings (Appendix E), so the number of buildings in a city grows sublinearly \cite{GrowthBettencourt, ScalingInteractions, bettencourt2013origins}. Put together, all the buildings' footprint comprises 10,000 constructed km2 (which is roughly the surface of Lebanon). The majority are small residential buildings (68.5\% of the buildings are less than 50 m2, and only 0.3\% have an area bigger than 500 m2). However, big buildings, although small in quantity, might represent a large part of the constructed surface of a city (Appendix F). Buildings in African cities have an average surface of 55 m2. However, the mean area of buildings also varies with city size. Roughly 18\% of the constructed surface of Africa is buildings larger than 250 m2. In Abidjan, for example, only 5\% of the buildings are bigger than 250 m2, but they total 30\% of the constructed surface of the city (Figure \ref{MsFigure1}-D). The average size of buildings scale with its population. A city with ten times the population has buildings that are, on average, 17\% bigger, the result of a disproportionate presence of big buildings in larger cities (Appendix F). The same applies across the whole continent. In West Africa, for instance, a city with ten times the population has buildings that are 40\% bigger. Large cities in the US and OECD countries are denser \cite{louf2014congestion}, but this is only observed in North Africa. In the rest of African regions, the footprint of a city increases superlinearly since cities have fewer buildings per person, but those buildings tend to be bigger. In West Africa, for instance, a city with ten times the population has 12.5 times the footprint (Appendix E). Further, larger cities have slightly taller buildings with a higher volume per person than smaller cities, mostly since the footprint is larger (Appendix H).

Many cities are elongated, especially urban areas dominated by small buildings, close to physical barriers, international borders or more polycentric (Figure \ref{MsFigure1}). However, as the population increases, distances grow and become critical, so cities experience intense competition for space and make better use of it, which results in less elongated cities \cite{batty2008size}. The mean distance between buildings is, on average, 2.8 times larger because of the non-circular shape of African cities. However, in West, South and Central Africa, larger cities are more compact (smaller sprawl $S$) and tend to be less elongated (smaller $E$), so even if they occupy more space, they settle in a more efficient manner (Appendix E). The street network is also not uniform across the continent. More elongated and sprawled cities have street networks with longer streets, even after controlling for population size and topography (Appendix C). At a regional level, North African cities have the lowest average street lengths across the continent (82 m), less than half of Central African cities (211 m), making them more walkable and liveable.

If large cities were just a scaled version of small cities, the mean distance between buildings should grow with the square root of their population. However, distances in large cities grow slightly faster, with exponent $\beta_D = 0.532$. Also, results range between $\hat{\beta}_C = 0.400$ for cities in Central Africa to $\hat{\beta}_N = 0.574$ for cities in North Africa. Even ignoring that cities expand vertically when the population approaches approximately 100,000 people \cite{molinero2021geometry}, it is observed that in North Africa, people occupy more space at an individual level in large cities, increasing its negative land use consequences \cite{foley2005global}. That extra space results from fewer but larger buildings and more elongated and sprawled cities. The same is not true across the rest of the continent. In West Africa, for example, larger cities are less elongated and have a smaller sprawl. It slightly counters the growth of distances due to a higher number of larger buildings in large metropolitan areas. Cities face two opposing forces as they grow. People demand more infrastructure, so more and larger buildings. But also, cities tend to be rounder and more compact to reduce distances and efficiently use space (Appendix E).

\begin{figure*} \centering
\includegraphics[width=\textwidth]{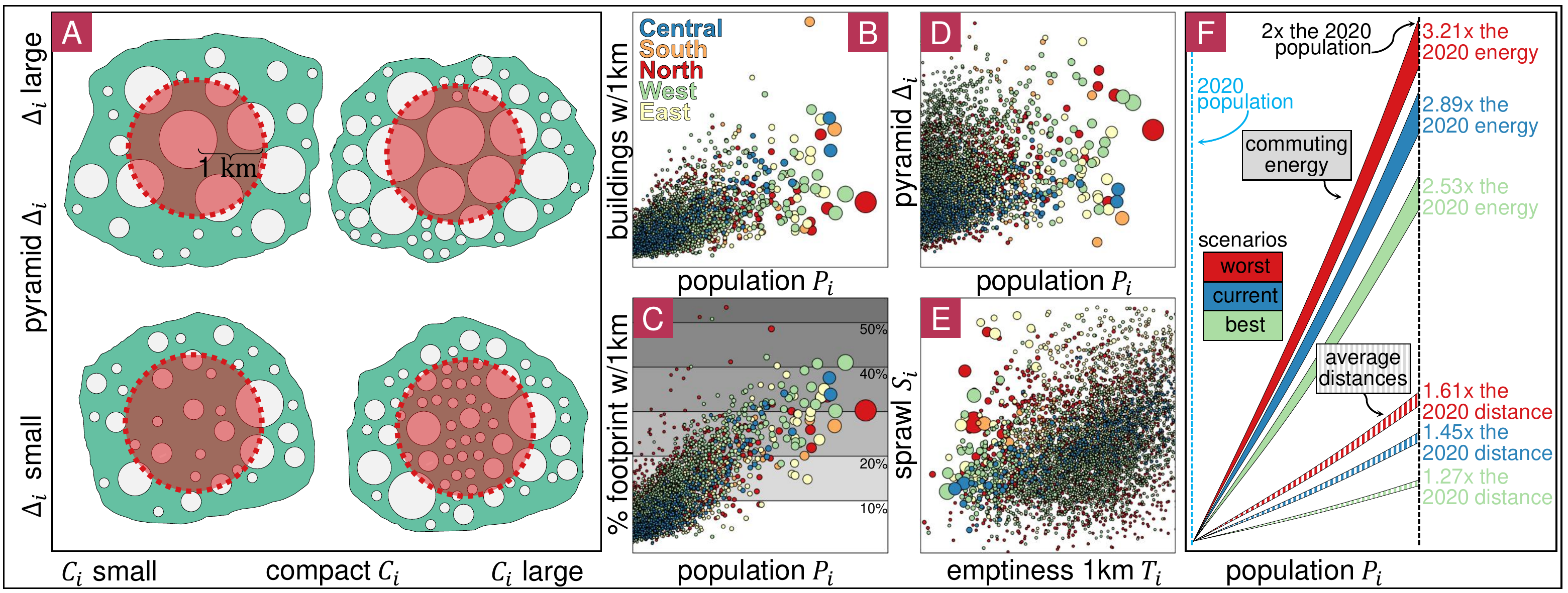}
\caption{A - The centre of a city, identified as its most dense location, and its 1 km vicinity is formed by some buildings that may be bigger than the rest of the city (forming a pyramid city) or even smaller (forming a valley), and the vicinity might have a small or large footprint. It is possible to characterise cities by looking only at the city centre. B, C, D - More populous cities have more and bigger buildings by their centre, so the footprint near the centre is also bigger. Among cities of more than one million inhabitants, at least 30\% of its centre is built. E - Most cities (except for some cities mostly in East Africa) have less sprawl with lower levels of emptiness. Thus, efficient use of space near the city centre is crucial for reducing city sprawl. }
\label{MsFigure2}
\end{figure*}

It is possible to characterise cities by observing only their densest location in terms of buildings (Figure \ref{MsFigure2}). The densest point in a city with more than 1 million people has 30 or 40\% of the surface built-up, as opposed to small towns with less than 100,000 inhabitants, where less than 10\% has been constructed (Appendix G). Thus, the sprawl at the densest point in a city with ten times more people is 63\% smaller. The densest point of a city might be ``empty'', meaning that its constructed surface might be small, suggesting that the whole urban area has a very low density. The emptiness of the densest point increases its sprawl and the energy requirement of cities (Appendix G), which may result from a town having only a limited number of (mostly small) constructions. But city size is also related to the structure of a city by its centre. Buildings close to the densest point tend to have a bigger footprint in a larger city (Figure \ref{MsFigure2}). For example, a city with ten times more population has 22\% bigger buildings near the centre (Appendix H). Also, most cities expand vertically for their densification process \cite{angel2021anatomy}. Considering the total volume within the densest point across cities, we find that larger cities are more constructed and expand vertically with taller buildings. Larger cities have more footprint and are more vertical within their densest point. A city with ten times more population has 3.4 times more infrastructure volume nearby its densest location since they have more constructed surface and buildings that are 25\% taller (Appendix H).

It has been estimated that if current trends in urban expansion continue, urban energy use will reach more than 730 exajoules (EJ) by 2050 \cite{creutzig2015global}. Our proxy for the change in the energy that the city requires for its mobility $T_i$ is the product of its population and mean distances, so $T_i \propto P_i D_i \propto P_i^{1.532}$ for the continent. Thus, a city with ten times the population requires 34 times more energy for mobility since more people commute even longer distances (Appendix I). Although small cities often lack urban expansion strategies, it is in large cities where the burden of sprawl and elongation is more significant. In the top 1\% largest cities in Africa (50 cities) is where 40\% of the population lives but where 80\% of the total energy consumed in intercity transportation occurs.

Some African countries, such as Niger or Chad, will double their 2020 population before 2050 \cite{Tusting2019}. Due to the urbanisation process and population growth, African cities will keep growing at an unprecedented speed, and some might reach a population of 80 or even 100 million inhabitants \cite{100MillionCities}. Assuming the population is the most important determinant for cities, we analyse the evolution of its form as cities grow \cite{pumain2004scaling, depersin2018global}. With a population of 80 million inhabitants, a city could have 18 million buildings (roughly two-thirds of the number of buildings there are today in Nigeria) with a footprint of 175 km2 and an average distance between buildings above 85 km (Appendix I). The burden on city dwellers of such an urbanisation process could force the city to grow in a roughly circular, compact and more vertical shape, urbanising green and blue areas. When a city grows, more people suffer longer distances, requiring even more energy in transport. We analyse what will happen for a city that doubles its 2020 population in terms of its footprint, distances and energy required for transport (Figure \ref{MsFigure2}-F). By the time African cities double their 2020 population, the average distance between buildings will increase 45\%, but more people will experience longer distances, so the energy consumed in transport in those cities will be three times the current levels (Appendix I). A city that doubles its population will experience at least 27\% longer distances (up to an increase of 61\% in the worst-case scenario, with high elongation and sprawl) and will require up to 3.21 times more energy in transport (Appendix I). Thus, some African cities rely on reducing the number of daily journeys and spatial mismatch and on fast mass transport to avoid collapsing for excessive congestion, and energy demands \cite{ewing1997angeles, creutzig2015global}.

\section{Discussion}

The richness of buildings' data enables the characterisation of African cities' urban morphology at a level of granularity that has never before been possible, using the novel BASE model. It is feasible to characterise cities' urban morphology solely on the building size and footprint at the densest point of a city (Appendix H). Such advancements are imperative so cities can guide urbanisation in the forthcoming decades towards a pathway of resilience and sustainability. Our results show that future energy needs for transport could be incredibly cumbersome if trends continue. The magnitude of these needs, along with how these needs are met (with or without fossil fuels), will set the future course of emissions, pollution, congestion, noise – and ultimately, the liveability of cities \cite{anderson2019natural}.

Accelerating urbanisation towards more compact, mixed land-use and polycentric morphologies is associated with greater sustainability and resilience, elsewhere in the world. Concerning transport, these urban forms have shorter blocks and more connected street networks, facilitating mobility and the utilisation of non-motorised modes, such as public transport and walking. However, steering urban growth in this form can place pressure on green spaces, diminishing the resilience of cities' to floods, heatwaves and landslides, along with losing other ecosystem services provided by these spaces like pollution and carbon absorption and biodiversity conservation. However, building taller buildings can reduce pressure on green spaces by accommodating the growing number of people. Urban morphology goes hand-in-hand with a city's future resilience and sustainability.  

This data is publically available for local actors on the OECD/SWAC's \href{https://mapping-africa-transformations.org/}{Mapping Africa Transformations Platform (MAPTA)} and updated over time with future releases of the raw data. Freely accessible data helps local actors in African cities to track the evolution of the built environment and guide decision-making, policies, funding, and regulations in these cities, in addition to fostering peer learning and shared experiences. Climate change is a global challenge, but mitigation and adaptation rest on the shoulders of local actors.

\section{Methods}

\subsection{Data construction}

\begin{figure*} \centering
\includegraphics[width=\textwidth]{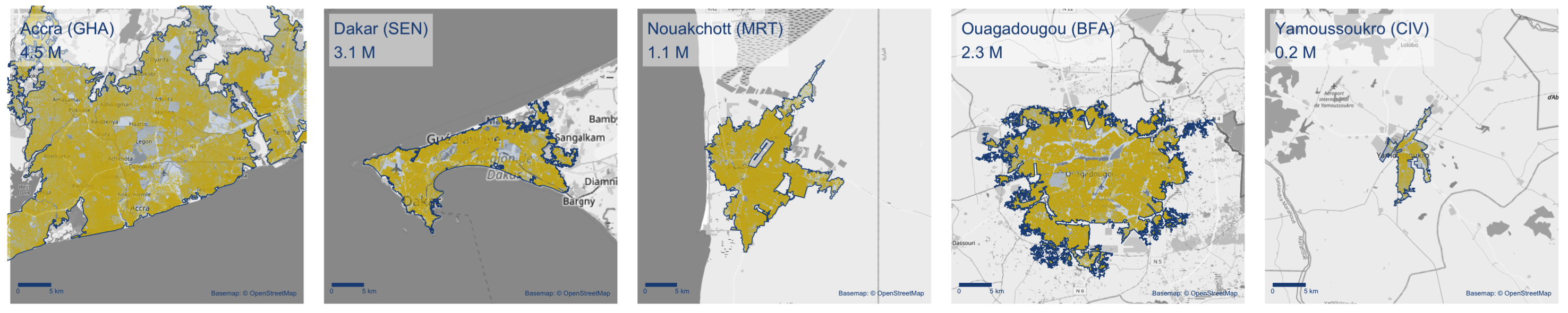}
\caption{Google Open Buildings locations over Africapolis urban agglomeration boundaries. Examples from 5 capital cities in West Africa: Accra (Ghana), Dakar (Senegal), Nouakchott (Mauritania), Ouagadougou (Burkina Faso), and Yamoussoukro (Ivory Coast). Maps at the same spatial scale.}
\label{fig:data_buildings_afpolis}
\end{figure*}

Defining metropolitan areas is challenging and often depends on considerations and parameters \cite{rozenfeld2008laws, cottineau2017diverse, arcaute2015constructing}. Africapolis applies the same definition for an urban agglomeration at a continental level (i.e., an agglomeration of at least 10,000 inhabitants and buildings less than 200 m apart), enabling it to analyse and compare cities between different countries \cite{Africapolis}. Buildings' locations were extracted from the recently launched Google Open Buildings dataset, \url{https://sites.research.google/open-buildings/}. The buildings' footprints were obtained using a deep learning model with high-resolution satellite imagery (50 cm pixel size) \cite{sirko2021continental}. We extract the buildings' footprints that are located within Africapolis polygons and keep the attributes of the building's centre location (latitude and longitude), the confidence score, and the building footprint area. We assign each building the unique identifier of the Africapolis urban agglomeration to which it belongs (Appendix A). This way, we obtained the location and attributes of the buildings' footprint of 6,849 African urban agglomerations (Figure \ref{fig:data_buildings_afpolis}). We computed street network metrics \cite{Boeing2017, OpenStreetMap} and terrain metrics \cite{NASAJPL2020}, such as the difference in elevation between the highest and lowest point, the average slope, and the average height within each Africapolis urban agglomeration polygon (Appendix A).

\subsection{The BASE model of cities}

For city $i$ with population size $P_i$, we compute the mean distance between pairs of buildings, $D_i$. Values of $D_i$ are larger due to four reasons: (1) city $i$ has more buildings, (2) its buildings are bigger, (3) buildings are arranged diffusely, and (4) the city has an elongated shape. Here we capture the four factors contributing to a city having longer distances and characterise them depending on distinct city attributes, such as differences in elevation and city size. Let $B_i$ be the number of buildings in the city $i$ and $A_i$ be their average size in m2. Let $S_i > 0$ be a coefficient for the diffusion of buildings, and $E_i\geq 1$ be a coefficient that captures how elongated is the shape of the city (Figure \ref{MsFigure1}). Cities mainly grow from the bottom up and adjust to the topography, barriers and road infrastructure, so in general, the shape of cities is not circular \cite{batty2008size}. Our data enables us to measure the mean distance $D_i$, the number of buildings $B_i$ and their size $A_i$. We construct $S_i$ and $E_i$ for each city. With values of $E_i$ close to one, the city's shape is nearly circular, and higher values represent more elongated shapes. With small values of $S_i$, the city has a small sprawl, so buildings are arranged compactly. Thus, with small $E_i$ and $S_i$, the city has roughly a circular shape and compacted buildings. The impact of $E_i$ and $S_i$ is to increase distances, taking a tight circle as the basis. We construct our BASE model based on the formula for the expected distance between two points inside a circle. The average distance between any two points inside a circle with area $a$ is given by $128\sqrt{a} /(45 \pi)$, so distances grow proportional to the radius. Equation \ref{MeanDist} is inspired by the average distance between points inside a circle, where instead of the radius, we use the city's footprint, $B_i A_i$, and two shape parameters with a multiplicative impact, $S_i E_i$.

Inspired by an ellipse, we measure the elongation of a city. Thus, with $E_i$, we capture how ``elliptical'' the city's shape is. The mean distance between two points inside an ellipse has no closed solution \cite{parry2000probability}, but an approximation can be constructed by considering the ratio between the major and the minor axis. Thus, we define the elongation $E_i$ as
\begin{equation} \label{EccEqn}
E_i = \frac{\sqrt{\pi} M_i}{2\sqrt{B_i A_i}},
\end{equation}
where $M_i$ is the ``diameter'' of the city, that is, the longest distance between any two buildings and $2\sqrt{B_i A_i / \pi}$ is the smallest possible diameter of a circle with $B_i A_i$ as a footprint. Smaller values of $E_i$ mean that the ratio between the smallest and the largest radius are similar, so the city's shape is more circular. Larger values mean more elongated shapes. In equation \ref{EccEqn}, $M_i$ is the ``major axis'', and $2\sqrt{B_i A_i / \pi}$ is the ``minor axis'' of a city (Appendix B). By considering $E_i$ to be the ratio between two distances, we obtain a scale-free coefficient $E_i \geq 1$ concerning the number or area of buildings. That means that if a city is a scaled version of another, for example, with four times the number of buildings (or buildings four times the size), then distances also grow, including doubling the maximum distance and doubling the smallest diameter, so $E_i$ remains the same.

We define the sprawl of a city as everything else that increases distances in cities besides the number of buildings, their area and elongation. From equation \ref{MeanDist} we get that
\begin{equation} \label{SprawlEquation}
S_i = \frac{45^2 \pi^{3/2} D_i^2}{2^{13} M_i \sqrt{B_i A_i}} = \gamma \frac{D_i^2}{M_i \sqrt{B_i A_i}},
\end{equation}
for $\gamma = 45^2 \pi^{3/2} / 2^{13} \approx 1.38$. 

Imagine that instead of buildings of a city, we have toy bricks of different sizes. There are infinitely many configurations to arrange those bricks keeping the maximum distance fixed (with a fixed diameter), meaning many brick configurations have the same elongation. In one extreme, the bricks ``fill'' the area with the corresponding major axis, but in the other extreme, the largest distance is observed only between a few bricks. Therefore, the sprawl $S_i$ captures those possible brick arrangements. 

Our technique decomposes the mean distance between buildings into four multiplicative components. Two components ($B_i$ and $A_i$) are measured directly from the data, and we have constructed a mathematical expression for $S_i$ and $E_i$, the sprawl and elongation. Thus, we have an exact expression that equates the mean distance between buildings in a city with four urban indicators (Equation \ref{MeanDistCircleUS}). 

\begin{equation} \label{MeanDistCircleUS}
D_i = \frac{128}{45\pi} \left( \underbrace{B_i A_i}_\text{footprint} \underbrace{S_i E_i}_{\text{shape}} \right)^{(1/2)}.
\end{equation}

One way to interpret the elongation and the sprawl of a city is that the mean distance between buildings in cities increases proportionally to $\sqrt{E_i}$, so if a city has an elongation value of $E_i = 4$, then the mean distance between buildings is double because the city is elongated, and similarly for the sprawl $S_i$.

\subsection{Minimum distances and fragmentation}

The number of buildings in a city is the most obvious reason why distances grow in large cities. More people means more buildings which then translates into larger distances. The same occurs with building size, so we ``discount'' those two reasons. Equation \ref{MeanDist} enables us to detect how small distances could be if the city was compact and had a circular shape, with area $B_i A_i$ and with $E_i = 1$ and $S_i = 1$. The smallest mean distance $D^\star_i$ is given by
\begin{equation} \label{OptDist}
D^\star_i = \frac{128}{45 \pi} \sqrt{ B_i A_i}.
\end{equation}
We construct the fragmentation index $\psi_i$ of a city, by
\begin{equation} \label{FragmentationEqn}
\psi_i = \frac{D_i}{D^\star_i} = \sqrt{S_i E_i},
\end{equation}
that indicates how distances in city $i$ are longer due to an elongated shape and more spacing between buildings. Fragmentation is a coefficient $\psi_i \geq 1$ independent of the number and area of buildings. 

\subsection{Measuring polycentrism}

Cities are increasingly characterised by polycentricity, that is, the presence of multiple interconnected centres \cite{thomas2021toward}. Measuring polycentrism is based on three methodological stages: delineating urban regions, identifying subcentres and then applying some mathematical function to obtain an index for a city \cite{thomas2021toward}. Different techniques have been used based on infrastructure data, such as satellite images, the density of points of interest, road network or buildings data \cite{xie2019modeling, korah2019characterising, deng2019detecting, taubenbock2017measuring} and also based on mobility data or travel surveys \cite{limtanakool2007theoretical, RevealingCentrality, toole2012inferring}. Measuring polycentrism depends not only on the type of data and the criteria applied for defining centres and subcentres but also on the possibly divergent methods of measuring polycentricity \cite{bartosiewicz2020investigating}.

Here, footprint data of millions of buildings is used to identify the spatial clustering of buildings in cities. The kernel density estimates the intensity of the number of buildings and the constructed surface per unit area \cite{Baddeley}. For example, the technique has been used for constructing a continuous surface representing the intensity of human activity based on points of interest in a city \cite{deng2019detecting}. The kernel density is a cumulative function obtained by adding a decaying surface for each building. Formally, for some point $x$ in space, we define the kernel as
\begin{equation}
k(x) = \sum_{j = 1}^n w_j f(d_{x,j}),
\end{equation}
where $n$ is the number of buildings, $w_j$ are the weights of each building, taken to be its area, and $d_{x,j}$ is the distance between the point $x$ and the centroid of the building, and $f$ is a decreasing function, known as the smoothing kernel, here taken to be a Gaussian function. The result is a surface over each urban area that highlights parts with more or larger constructions (peaks) considered centres of the city \cite{Diggle}.

Based on the kernel density, a contour adjacent relation tree is constructed \cite{li2018extraction}, where each node on the tree is a new contour. Contour trees represent spatial relationships between the contours of the kernel surface and summarise relationships at different levels \cite{guilbert2013multi}. Separate peaks (urban centres) are identified as branches on a tree, which are connected depending on the different contour levels of the surface. The procedure gives $N$ branches, where $N=1$ is a monocentric city. Each branch has three indicators: height (corresponding to the kernel estimate), area (representing the total surface of the city that belongs to that branch) and volume (obtained by multiplying the area and height of each branch). 

Let $Br_i$ be the number of branches of city $i$ and let $v_k$ be the volume of each branch in decreasing order (so that $v_1 \geq v_2 \geq \dots \geq v_{Br_i}$). The polycentrism index $\phi_i$ is defined as 
\begin{equation} \label{PolycentrismFormula}
\phi_i = \frac{1}{v_1} \sum_{k = 1}^{Br_i} k v_k.
\end{equation}

\begin{figure} \centering
\includegraphics[width=0.4\textwidth]{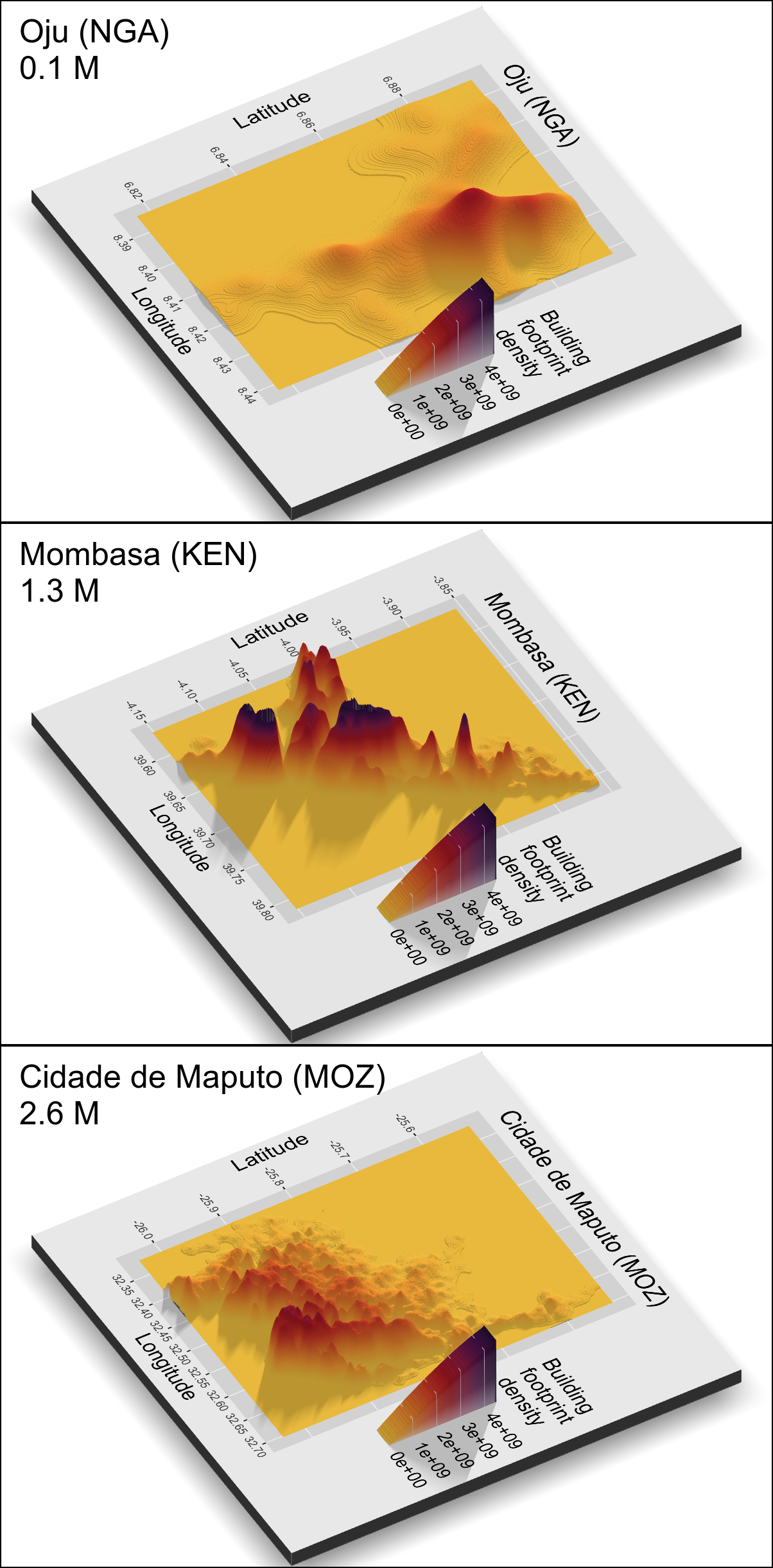}
\caption{Three-dimensional representation of buildings' footprint density. \textit{Top:} Oju (Nigeria), a monocentric city with low levels of elongation and sprawl; \textit{middle:} Mombasa (Kenya), a polycentric city high levels of elongation but low levels of sprawl; \textit{bottom:} Maputo (Mozambique), a polycentric city with a low level of elongation but a high level of sprawl. Each city plotted at a different spatial scale (3D plots created using \textit{rayshader} package \cite{Morgan-Wall2021} in R).}
\label{fig:density_3D}
\end{figure}

The index $k$ inside the sum of equation \ref{PolycentrismFormula} helps increase the value of $\phi_i$ with the number of branches. By dividing by $v_1$ in equation \ref{PolycentrismFormula}, a comparable index across cities is obtained (Appendix D). If $\phi_i = 1$ then $i$ is monocentric. A city with two distant and equal-sized centres (so that they belong to different branches) has $\phi_i = 3$ (with three equal-sized centres $\psi_i =6$, and so on). The procedure also gives the volume tree of a city, where the dimension of each branch (horizontal axis) and its height (vertical axis) represent the centres of the city \cite{klemela2009smoothing, li2018extraction}. Larger cities tend to be more polycentric, formed by more branches with more separation between them and a larger relative volume. However, medium-size cities may also be highly polycentric, reflecting fragmented urban areas (Figure \ref{fig:density_3D}). 

\subsection{Impact of city size}

The infrastructure and the socio-economic outputs of a city vary according to many factors, and city size has been detected to be a critical aspect of cities \cite{pumain2004scaling, GrowthBettencourt, ScalingInteractions}. We characterise the four city indicators of the BASE model and detect if they vary according to city size. We fit equations like $B_i = \alpha P_i^{\beta}$ between the number of buildings $B_i$ and the population $P_i$ for some coefficients $\alpha$ and $\beta$, and similarly for the other indicators. Values of $\beta \approx 0$ indicate that city size has little or no impact on the corresponding indicator. Values of $\beta \approx 1$ indicate a linear growth, and values below (and above) $\beta = 1$ are a sublinear (or superlinear) impact of city size (Appendix E).

The scaling coefficients show a sublinear relation with the number of buildings $B_i$ (so fewer buildings per person as cities grow), sublinear with the area of buildings $A_i$ (so bigger buildings in larger cities). Also, larger cities tend to have slightly more sprawl, and the elongation is statistically the same across cities of different sizes (Table \ref{TableScalingModel}).

\begin{table}
\centering
\begin{tabular}{ c|cc|cc}
variable & \multicolumn{2}{|c|}{$\alpha$}  & \multicolumn{2}{c}{$\beta$}\\
\hline
$B_i$ - number of buildings & 0.423 & 0.029 & 0.981 & (0.007) \\
$A_i$ - area of buildings & 24.85 & 1.29 & 0.067 & (0.005) \\
$S_i$ - sprawl & 0.725 &  0.067 & 0.011 & (0.006) \\
$E_i$ - elongation & 5.925 & 0.337 & 0.005 & (0.005)\\
\hline
$D_i$ - mean distance & 6.126 & 0.479 & 0.532 & (0.006)\\
\end{tabular}
\caption{Observed scaling coefficients at the continental level}
\label{TableScalingModel}
\end{table}

Many historical and geographical differences across regions and countries in Africa also impact how cities have grown. Thus, the impact of city size is not uniform across the continent. For example, in West Africa, buildings are bigger in larger cities (so, a city with ten times the population has buildings that are 40\% bigger). In contrast, in South Africa, buildings tend to be the same size across all cities (Table \ref{TableScalingModel}). Cities in Central and South Africa become rounder and more compact as they grow, as opposed to cities in North Africa, where elongation and sprawl increase in larger cities.

\begin{table}
\begin{center}
{\small
\begin{tabular}{l c c c c c}
\hline
 & North & West & East & Central & South \\
\hline
$B_i$ & $0.926^{***}$ & $0.957^{***}$  & $1.042^{***}$  & $0.990^{***}$  & $1.005^{***}$ \\
                      & $(0.013)$     & $(0.011)$      & $(0.010)$      & $(0.017)$      & $(0.014)$     \\
$A_i$ & $0.044^{***}$   & $0.144^{***}$   & $0.035^{***}$   & $0.076^{***}$   & $0.018$         \\
                      & $(0.009)$       & $(0.010)$       & $(0.007)$       & $(0.012)$       & $(0.010)$       \\
$S_i$ & $0.106^{***}$  & $-0.026^{**}$ & $0.026^{*}$  & $-0.156^{***}$ & $-0.078^{***}$ \\
                      & $(0.012)$      & $(0.009)$     & $(0.013)$    & $(0.014)$      & $(0.019)$      \\
$E_i$ & $0.071^{***}$ & $-0.017^{*}$  & $-0.001$      & $-0.109^{***}$ & $-0.065^{***}$ \\
                      & $(0.010)$     & $(0.007)$     & $(0.010)$     & $(0.015)$      & $(0.019)$      \\
\hline
$D_i$ & $0.574^{***}$  & $0.528^{***}$  & $0.551^{***}$  & $0.400^{***}$  & $0.440^{***}$  \\
                      & $(0.012)$      & $(0.008)$      & $(0.011)$      & $(0.012)$      & $(0.022)$      \\
$\psi_i$ & $0.089^{***}$ & $-0.022^{**}$ & $0.013$       & $-0.133^{***}$ & $-0.072^{***}$ \\
                      & $(0.010)$     & $(0.008)$     & $(0.011)$     & $(0.013)$      & $(0.018)$      \\
$\phi_i$ & $0.310^{***}$  & $0.186^{***}$  & $0.325^{***}$  & $0.103^{***}$  & $0.259^{***}$  \\
                      & $(0.009)$      & $(0.006)$      & $(0.010)$      & $(0.008)$      & $(0.014)$      \\
$A100_i$ & $0.078^{***}$  & $0.201^{***}$  & $0.122^{***}$  & $0.204^{***}$  & $0.048^{**}$   \\
                      & $(0.014)$      & $(0.013)$      & $(0.013)$      & $(0.024)$      & $(0.017)$      \\
$\theta_i$ & $0.970^{***}$ & $1.100^{***}$ & $1.078^{***}$ & $1.066^{***}$ & $1.023^{***}$ \\
                      & $(0.014)$     & $(0.009)$     & $(0.011)$     & $(0.018)$     & $(0.018)$     \\
$\theta 1km_i$ & $0.408^{***}$ & $0.471^{***}$ & $0.413^{***}$ & $0.515^{***}$ & $0.402^{***}$ \\
                      & $(0.013)$     & $(0.010)$     & $(0.012)$     & $(0.019)$     & $(0.022)$     \\          
\hline
\multicolumn{6}{l}{\scriptsize{$^{***}p<0.001$; $^{**}p<0.01$; $^{*}p<0.05$}}
\end{tabular}
}
\caption{Observed scaling coefficients across regions in Africa}
\label{RegionalTablesBeta}
\end{center}
\end{table}

The mean distance between buildings is also affected by city size. Using equation \ref{MeanDist} to compute the scaling coefficient between buildings is also possible. Since $D_i = \frac{128}{45 \pi} \sqrt{ B_i A_i S_i E_i}$, then we can write $D_i$ as it varies with city size and obtain that $D_i = \gamma P_i^{(\beta_B +\beta_A + \beta_S + \beta_E)/2}$, with $\gamma =\frac{128}{45 \pi} \sqrt{\alpha_B \alpha_A \alpha_S \alpha_E}$. Thus, the scaling coefficient of distances is obtained by half the sum of the scaling of buildings, area, sprawl and elongation (Appendix I). For example, in North Africa, the scaling coefficient of the distance is $_N\beta_D = (0.926 + 0.044 + 0.106 + 0.071)/2 = 0.574$ so the mean distance between buildings in cities increase at a faster rate than the square root of the population. Having decomposed the mean distance into four urban components enables us to detect that in North Africa, the reasons why distances are longer in larger cities are first the number of buildings, then the increasing sprawl and elongation and, to a small extent, having bigger buildings. But in South Africa, having more and bigger buildings in cities increases distances in larger cities, but a decreasing sprawl and elongation contribute to reducing distances. In South Africa, the mean distance between buildings in cities increases at a slower rate than the square root of the population, with $_S\beta_D = 0.440$. Suppose larger cities were just a scaled version of a small city. In that case, the number of buildings per person should be constant (so $\beta_B = 1$), and the rest of the scaling coefficients should be zero, reflecting the same building size, sprawl and elongation, so that the mean distance should grow with the square root of the population. However, cities are not a scaled version of a small city, and most components vary with city size.

Beyond distances, it is possible to decompose a city's footprint and fragmentation. The footprint $\theta_i$ can be decomposed as $\theta_i = B_i A_i = \alpha_{\theta} P^{\beta_B + \beta_A}$, so for example, the footprint is superlinear in West Africa but sublinear in North Africa. The fragmentation $\psi_i$ can be decomposed as $ \sqrt{S_i E_i} = \alpha_{\psi} P_i^{(\beta_S + \beta_E)/2}$, where $\alpha_{\psi} = \sqrt{\alpha_S \alpha_E}$ and where the population has a scaling coefficient $(\beta_S + \beta_E)/2$. Thus, larger cities in North Africa are more fragmented than smaller cities mainly due to a higher sprawl, but the opposite happens in Central Africa, where the sprawl and elongation are smaller in larger cities, so they tend to be less fragmented (Appendix E). 

\subsection{Characterising cities based on the city centre}

Considering the distribution of the urban footprint at a distance $R=1$ km from its centre enables us to observe the sprawl across cities based on the same shape. Comparing the same shape across different urban polygons means we can ignore the elongation and focus only on the sprawl. The sprawl of a city inside a circle, referred to as the emptiness $T_i^{(R)} \geq 1$, is defined as the ratio between the surface of the city centre and the part that is constructed, given by
\begin{equation}
T_i^{(R)} = \frac{\pi R^2}{B_i^{(R)} A_i^{(R)}},
\end{equation}
where $B_i^{(R)}$ and $A_i^{(R)}$ are the number and average area of buildings inside that circle. The emptiness $T_i^{(R)}$ corresponds to the observed sprawl inside a circle, and a higher emptiness means that a smaller surface of the centre has been constructed (Appendix G). Although longer distances than $R = 1$ km may also be considered for observing cities at their centre, we observe that a circle with a 1-km radius is the largest circle that fits almost always inside the polygon of all cities. Bigger circles often are not fully contained inside the polygon of cities, so analysing the footprint depends on some elongation, whereas smaller circles have a smaller building sample in cities.

\section{Appendix}

\subsection{Data construction} 

We use open geospatial data sets for this analysis: urban boundaries, buildings' locations, street network data, and elevation data. The boundaries of African urban agglomerations were downloaded from the Africapolis website in November 2021 \cite{Africapolis}. The downloaded dataset contains 7720 features with polygon geometry representing urban agglomeration boundaries by 2015, packed in ESRI Shapefile format. Defining metropolitan areas is challenging and often depends on certain considerations and parameters. Africapolis applies the same definition for a metropolitan area, enabling us to compare cities at a continental level. According to the Africapolis website, an urban agglomeration unit combines satellite and aerial imagery and official demographic data such as censuses and other cartographic sources. The delineation of an urban agglomeration boundary is based on a spatial approach that combines physical limits, the continuity of the built-up area, and demographic criteria to provide a standardised continental spatial definition of urban areas. All urban agglomerations considered have more than 10,000 inhabitants. 

Buildings' locations were extracted from the recently launched Google Open Buildings dataset, \url{https://sites.research.google/open-buildings/} which contains 516M building footprints across an area of 19.4M km$^{2}$, 64\% of the African continent \cite{sirko2021continental}. The buildings' footprints were obtained using a deep learning model with high-resolution satellite imagery (50 cm pixel size) \cite{sirko2021continental}. This dataset is packed into 136 different files in comma-separated values format. It contains information on the location of the building's footprint centre in geographic coordinates (latitude and longitude), its area in m2, a \textit{Plus Code} corresponding to the centre of the building, and a confidence score that informs about the confidence level of the building detection, and the geometry of the building's footprint. See \url{https://maps.google.com/pluscodes/} for more information on Plus Codes.

We downloaded the 136 Open Buildings files (49.8 GB) and implemented a geoprocessing workflow in R software \cite{RCoreTeam2018} to extract the buildings' footprints within Africapolis polygons. We keep the attributes of the building's centre location (latitude and longitude), the confidence score, and the building footprint area. We assign the unique identifier (agglosID) of the Africapolis urban agglomeration it belongs. This way, we obtained the location and attributes of the buildings' footprints of 6,849 African urban agglomerations. There are 871 Africapolis urban agglomerations without Google Open Building data. In Morocco, Western Sahara, Libya, Chad, Cameroon, and South Sudan, as well as North Kivu Province (Democratic Republic of Congo), Cabo Delgado Province (Mozambique), and the Southwestern region in Sudan, all lack buildings' footprints in the Google Open Buildings dataset. We also combine the buildings' footprint with data provided by the German Aerospace Centre, giving the average height and volume of different regions of the city \cite{esch2022world}.

We computed street network metrics (the average street length and the total number of nodes or intersections) using the Python library OSMnx \cite{Boeing2017} with the Africapolis urban agglomeration polygons. We used that information to compute the number of buildings per street node for each urban agglomeration. We also calculated some terrain metrics, such as the difference in elevation between the highest and lowest point, the average slope, and the average height within each Africapolis urban agglomeration polygon using the NASADEM digital elevation model in Google Earth Engine \cite{NASAJPL2020}.

Since our procedure combines different geographical data, other filters are applied to the cities:

$\circ$ Drop 1,202 cities with less than 2,000 buildings since there is not enough data in the urban area.

$\circ$ Drop 123 cities with elongation $E_i >20$ and seven cities with sprawl $S_i >4$, reflecting that polygons and buildings do not spatially align. The mean elongation is $\mu_E = 7.5$ with $\sigma_E = 4.5$, and the mean sprawl is $\mu_S = 0.94$ with $\sigma_S = 0.83$, so cities with a higher elongation or sprawl are outliers.

$\circ$ Drop 24 cities with a footprint per person above 60 m2. The mean footprint per person is $\mu_\theta = 19$ m2 per person, with $\sigma_\theta = 20$.

\subsection{Measuring the elongation and sprawl of a city} 

\begin{figure*} \centering
\includegraphics[width=0.95\textwidth]{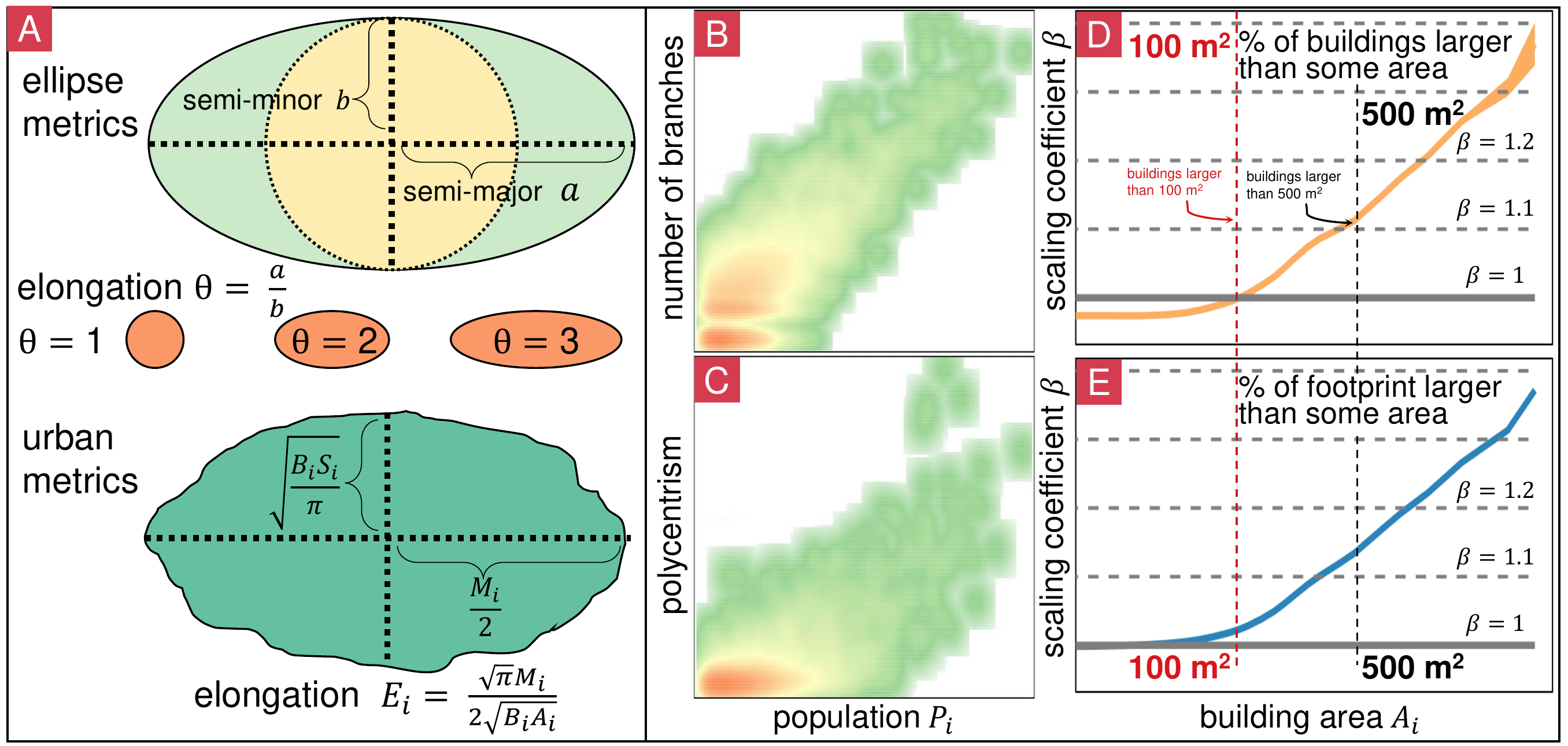}
\caption{A - The (modified) eccentricity of an ellipse $e'$ is defined as the ratio between the semi-major and semi-minor axis, both squared. Similarly, we define the city eccentricity $\theta_i$ as the ratio between the maximum observed distance (taken as the major axis) and the minimum radius of a circle with $B_i$ buildings with an average surface $A_i$. B, C - Number of branches (top) and polycentrism (bottom) observed in a city according to its population (horizontal axis). D, E - Scaling coefficient $\beta$ (vertical) of the number of buildings larger than some area (top) for different areas (horizontal axis). Scaling coefficient $\beta$ (vertical) of the area in the city constructed based on buildings larger than some area (top) for different areas (horizontal axis). }
\label{ApFigure}
\end{figure*}

Based on the eccentricity of an ellipse, we construct a metric using the ratio between the ``major'' and ``minor'' axis of a city. The major axis of a city is $M_i$, the maximum observed distance between buildings. If all city buildings were arranged next to the other in a circular shape, they would form a circle with area $B_i A_i$ and with radius $\sqrt{B_i A_i / \pi}$, considered to be the minor axis of a city. There is no closed formula to measure the mean distance between two points inside an ellipse \cite{parry2000probability}. A good approximation is given by $128b\theta/(45\pi)$, where $\theta = \sqrt{a/b}$, and where $a$ is the major axis $a$ and $b$ is the minor axis of the ellipse, so $\theta$ captures the discrepancy between $a$ and $b$. The formula is similar to the distance between two points inside a circle of radius $b$ but multiplied by a factor $\theta \geq 1$ that captures the shape of the ellipse. The formula is exact for $\theta = 1$ and works well for small values of $\theta$. Therefore, in an ellipse, distances grow roughly proportional to $\theta$. For city $i$, we construct its elongation $E_i$ as the ratio between the maximum observed distance between buildings and the minimum possible radius of a city. As with an ellipse, values of $E_i$ close to one mean a more circular shape, whereas higher values mean elongated cities across some axis. Also, distances in the city are expected to grow proportional to $\sqrt{E_i}$, so the expression aligns with the $BASE$ model (Figure \ref{ApFigure}-A).

Results show a scaling impact of city size that increases the number of buildings, their area, and the city's sprawl and elongation (Table \ref{AppTableBASE}). 

\begin{table}
{\tiny
\begin{center}
\begin{tabular}{l c c c c c c}
\hline
 & $y = \log B$ & $y = \log A$ & $y = \log S$ & $y = \log E$ & $y = \log D$ & $y = \log \theta$ \\
\hline
(Intercept)           & $-0.846^{***}$ & $-10.603^{***}$ & $-0.322^{***}$ & $1.779^{***}$ & $-5.095^{***}$ & $2.366^{***}$ \\
                      & $(0.068)$      & $(0.052)$       & $(0.063)$      & $(0.054)$     & $(0.061)$      & $(0.067)$     \\
log(Population\_2015) & $0.981^{***}$  & $0.067^{***}$   & $0.011$        & $0.005$       & $0.532^{***}$  & $1.048^{***}$ \\
                      & $(0.007)$      & $(0.005)$       & $(0.006)$      & $(0.005)$     & $(0.006)$      & $(0.006)$     \\
\hline
R$^2$                 & $0.799$        & $0.031$         & $0.001$        & $0.000$       & $0.594$        & $0.825$       \\
Adj. R$^2$            & $0.799$        & $0.031$         & $0.000$        & $-0.000$      & $0.594$        & $0.825$       \\
Num. obs.             & $5625$         & $5625$          & $5625$         & $5625$        & $5625$         & $5625$        \\
\hline
\multicolumn{7}{l}{\scriptsize{$^{***}p<0.001$; $^{**}p<0.01$; $^{*}p<0.05$}}
\end{tabular}
\caption{Scaling of the $BASE$ components, distance $D$ and of footprint $\theta$.}
\label{AppTableBASE}
\end{center} }
\end{table}

Further, we detect that some urban indicators beyond population also alter the urban footprint and shape, including the number of buildings (Table \ref{TableBuildings}), area (Table \ref{TableArea}), sprawl (Table \ref{TableSprawl}) and elongation (Table \ref{TableElongation}) of cities.

\begin{table}
{\tiny
\begin{center}
\begin{tabular}{l c}
\hline
 & $y = \log B$ \\
\hline
(Intercept)              & $2.4331 \; (0.0979)^{***}$  \\
log(Population\_2015)    & $0.5675 \; (0.0089)^{***}$  \\
RegionEast               & $0.0821 \; (0.0182)^{***}$  \\
RegionNorth              & $0.0655 \; (0.0211)^{**}$   \\
RegionSouth              & $0.5718 \; (0.0230)^{***}$  \\
RegionWest               & $0.1285 \; (0.0192)^{***}$  \\
log(Polycentrism)        & $0.4115 \; (0.0150)^{***}$  \\
log(dif\_elev\_m)        & $0.1140 \; (0.0057)^{***}$  \\
a100                     & $0.1232 \; (0.0586)^{*}$    \\
TotalFootprintCentre1km  & $0.0000 \; (0.0000)^{***}$  \\
MeanArea1000m            & $-0.0144 \; (0.0005)^{***}$ \\
log(street\_length\_avg) & $-0.0433 \; (0.0093)^{***}$ \\
IsPyramid $\Delta$       & $0.6257 \; (0.0231)^{***}$  \\
\hline
R$^2$                    & $0.9175$                    \\
Adj. R$^2$               & $0.9173$                    \\
Num. obs.                & $5625$                      \\
\hline
\multicolumn{2}{l}{\scriptsize{$^{***}p<0.001$; $^{**}p<0.01$; $^{*}p<0.05$}}
\end{tabular}
\caption{Model number of buildings $B$ depending of the population, region, polycentrism, differences in elevation, the area covered by buildings with more than 100 m2 (a100), the footprint within the centre and the area of buildings, the average street length and the ratio between the average area of buildings within 1 km of the city centre and the average area of buildings in the city (IsPyramid $\Delta$).}
\label{TableBuildings}
\end{center} }
\end{table}

\begin{table}
{\tiny
\begin{center}
\begin{tabular}{l c}
\hline
 & $y = \log A$ \\
\hline
(Intercept)              & $-10.372 \; (0.026)^{***}$ \\
log(Population\_2015)    & $-0.008 \; (0.002)^{***}$  \\
RegionEast               & $0.001 \; (0.005)$         \\
RegionNorth              & $-0.058 \; (0.006)^{***}$  \\
RegionSouth              & $-0.032 \; (0.006)^{***}$  \\
RegionWest               & $-0.071 \; (0.005)^{***}$  \\
log(Polycentrism)        & $-0.004 \; (0.004)$        \\
log(dif\_elev\_m)        & $0.008 \; (0.002)^{***}$   \\
a100                     & $0.877 \; (0.016)^{***}$   \\
TotalFootprintCentre1km  & $0.000 \; (0.000)^{***}$   \\
MeanArea1000m            & $0.010 \; (0.000)^{***}$   \\
log(street\_length\_avg) & $0.009 \; (0.004)^{*}$     \\
log(BuildingsPerNode)    & $-0.005 \; (0.002)^{**}$   \\
IsPyramid $\Delta$       & $-0.370 \; (0.006)^{***}$  \\
\hline
R$^2$                    & $0.951$                    \\
Adj. R$^2$               & $0.951$                    \\
Num. obs.                & $5625$                     \\
\hline
\multicolumn{2}{l}{\scriptsize{$^{***}p<0.001$; $^{**}p<0.01$; $^{*}p<0.05$}}
\end{tabular}
\caption{Model building area $A$ depending of the population, region, polycentrism, differences in elevation, the area covered by buildings with more than 100 m2 (a100), the footprint within the centre and the area of buildings, the average street length and the ratio between the average area of buildings within 1 km of the city centre and the average area of buildings in the city (IsPyramid $\Delta$).}
\label{TableArea}
\end{center} }
\end{table}

\begin{table}
{\tiny
\begin{center}
\begin{tabular}{l c}
\hline
 & $y = \log S$ \\
\hline
(Intercept)              & $-2.510 \;    (0.099)^{***}$ \\
log(Population\_2015)    & $0.025 \;    (0.010)^{*}$    \\
RegionEast               & $0.013 \;    (0.017)$        \\
RegionNorth              & $0.064 \;    (0.020)^{**}$   \\
RegionSouth              & $0.091 \;    (0.022)^{***}$  \\
RegionWest               & $-0.034 \;    (0.018)$       \\
MeanBuildingArea         & $1521.843 \; (1073.992)$     \\
log(Polycentrism)        & $0.399 \;    (0.015)^{***}$  \\
log(dif\_elev\_m)        & $0.049 \;    (0.006)^{***}$  \\
a100                     & $0.039 \;    (0.063)$        \\
TotalFootprintCentre1km  & $-0.000 \;    (0.000)^{***}$ \\
MeanArea1000m            & $-0.002 \;    (0.001)^{**}$  \\
log(street\_length\_avg) & $0.295 \;    (0.013)^{***}$  \\
log(nodes)               & $0.114 \;    (0.006)^{***}$  \\
IsPyramid $\Delta$       & $0.223 \;    (0.033)^{***}$  \\
ICoast                   & $0.071 \;    (0.014)^{***}$  \\
IBorder                  & $0.058 \;    (0.017)^{***}$  \\
\hline
R$^2$                    & $0.578$                      \\
Adj. R$^2$               & $0.576$                      \\
Num. obs.                & $5625$                       \\
\hline
\multicolumn{2}{l}{\scriptsize{$^{***}p<0.001$; $^{**}p<0.01$; $^{*}p<0.05$}}
\end{tabular}
\caption{Model sprawl $S$ depending of the population, region, the mean building area, polycentrism, differences in elevation, the area covered by buildings with more than 100 m2 (a100), the footprint within the centre and the area of buildings, the average street length and the ratio between the average area of buildings within 1 km of the city centre and the average area of buildings in the city (IsPyramid $\Delta$), whether the city is coastal and whether the city is within an international border (less than 50 km).}
\label{TableSprawl}
\end{center} }
\end{table}

\begin{table}
{\tiny
\begin{center}
\begin{tabular}{l c}
\hline
 & $y = \log E$ \\
\hline
(Intercept)              & $-0.211 \;   (0.088)^{*}$   \\
log(Population\_2015)    & $0.022 \;   (0.008)^{**}$   \\
RegionEast               & $0.064 \;   (0.015)^{***}$  \\
RegionNorth              & $-0.072 \;   (0.018)^{***}$ \\
RegionSouth              & $-0.189 \;   (0.020)^{***}$ \\
RegionWest               & $-0.109 \;   (0.016)^{***}$ \\
MeanBuildingArea         & $907.518 \; (949.954)$      \\
log(Polycentrism)        & $0.218 \;   (0.013)^{***}$  \\
log(dif\_elev\_m)        & $0.078 \;   (0.005)^{***}$  \\
a100                     & $0.071 \;   (0.056)$        \\
TotalFootprintCentre1km  & $-0.000 \;   (0.000)^{***}$ \\
MeanArea1000m            & $-0.002 \;   (0.001)^{*}$   \\
log(street\_length\_avg) & $0.275 \;   (0.012)^{***}$  \\
log(nodes)               & $0.092 \;   (0.005)^{***}$  \\
IsPyramid $\Delta$       & $0.085 \;   (0.029)^{**}$   \\
ICoast                   & $0.050 \;   (0.012)^{***}$  \\
IBorder                  & $0.037 \;   (0.015)^{*}$    \\
\hline
R$^2$                    & $0.548$                     \\
Adj. R$^2$               & $0.546$                     \\
Num. obs.                & $5625$                      \\
\hline
\multicolumn{2}{l}{\scriptsize{$^{***}p<0.001$; $^{**}p<0.01$; $^{*}p<0.05$}}
\end{tabular}
\caption{Model elongation $E$ depending of the population, region, the mean building area, polycentrism, differences in elevation, the area covered by buildings with more than 100 m2 (a100), the footprint within the centre and the area of buildings, the average street length and the ratio between the average area of buildings within 1 km of the city centre and the average area of buildings in the city (IsPyramid $\Delta$), whether the city is coastal and whether the city is within an international border (less than 50 km).}
\label{TableElongation}
\end{center} }
\end{table}

\subsection{Terrain, elevation and physical factors} 

The growth of cities is often bounded by physical barriers, such as mountains or coasts. Thus, when we observe that a city is very elongated, it is frequently because of rugged terrain and differences in elevation or because it is located beside a water body (river, lake, or sea). To measure the impact of physical barriers and rough terrain on the elongation of a city, we measure the distance to the nearest coast, the point with the lowest and the highest altitude in the city, and differences in altitude. This set of variables enables us to correlate the city metrics and detect whether being close to a coast increases the urban form indicators. Cities with high elevation or differences in altitude suggest proximity to mountains and the presence of physical barriers and rugged terrain.

Results show that larger differences in elevation and being close to a coastline are associated with higher levels of elongation (Table \ref{TableElongation}) and sprawl (Table \ref{TableSprawl}), and therefore higher levels of fragmentation (Table \ref{TableFragmentation}) and of polycentrism (Table \ref{TablePolcentrism}). Our models show that an increase of 1\% in elevation difference is associated with increases of 5\% in sprawl and almost 8\% in elongation. Being along a coastline is associated with 7\% more sprawl and 5\% more elongation. However, small cities along main roads also feature very high levels of elongation, even when the differences in elevation are not large. The most elongated cities in Africa are always located along a coastline or main roads.

\begin{table}
{\tiny
\begin{center}
\begin{tabular}{l c}
\hline
 & $y = \log \psi$ \\
\hline
(Intercept)              & $-1.360 \;   (0.081)^{***}$ \\
log(Population\_2015)    & $0.023 \;   (0.008)^{**}$   \\
RegionEast               & $0.039 \;   (0.014)^{**}$   \\
RegionNorth              & $-0.004 \;   (0.017)$       \\
RegionSouth              & $-0.049 \;   (0.018)^{**}$  \\
RegionWest               & $-0.072 \;   (0.015)^{***}$ \\
MeanBuildingArea         & $1214.681 \; (878.774)$     \\
log(dif\_elev\_m)        & $0.063 \;   (0.005)^{***}$  \\
a100                     & $0.055 \;   (0.052)$        \\
TotalFootprintCentre1km  & $-0.000 \;   (0.000)^{***}$ \\
MeanArea1000m            & $-0.002 \;   (0.001)^{**}$  \\
log(Polycentrism)        & $0.308 \;   (0.012)^{***}$  \\
log(street\_length\_avg) & $0.285 \;   (0.011)^{***}$  \\
log(nodes)               & $0.103 \;   (0.005)^{***}$  \\
IsPyramid $\Delta$       & $0.154 \;   (0.027)^{***}$  \\
ICoast                   & $0.061 \;   (0.011)^{***}$  \\
IBorder                  & $0.048 \;   (0.014)^{***}$  \\
\hline
R$^2$                    & $0.622$                     \\
Adj. R$^2$               & $0.620$                     \\
Num. obs.                & $5625$                      \\
\hline
\multicolumn{2}{l}{\scriptsize{$^{***}p<0.001$; $^{**}p<0.01$; $^{*}p<0.05$}}
\end{tabular}
\caption{Model fragmentation $\psi$ depending of the population, region, the mean building area, polycentrism, differences in elevation, the area covered by buildings with more than 100 m2 (a100), the footprint within the centre and the area of buildings, the average street length and the ratio between the average area of buildings within 1 km of the city centre and the average area of buildings in the city (IsPyramid $\Delta$), whether the city is coastal and whether the city is within an international border (less than 50 km).}
\label{TableFragmentation}
\end{center} }
\end{table}

\begin{table}
{\tiny
\begin{center}
\begin{tabular}{l c}
\hline
 & $y = \log \phi$ \\
\hline
(Intercept)              & $-4.004 \;   (0.074)^{***}$    \\
log(Population\_2015)    & $0.319 \;   (0.008)^{***}$     \\
RegionEast               & $0.163 \;   (0.016)^{***}$     \\
RegionNorth              & $0.330 \;   (0.018)^{***}$     \\
RegionSouth              & $0.195 \;   (0.020)^{***}$     \\
RegionWest               & $0.204 \;   (0.017)^{***}$     \\
MeanBuildingArea         & $-8458.903 \; (981.017)^{***}$ \\
TotalFootprintCentre1km  & $-0.000 \;   (0.000)^{***}$    \\
MeanArea1000m            & $0.009 \;   (0.001)^{***}$     \\
log(dif\_elev\_m)        & $0.027 \;   (0.005)^{***}$     \\
a100                     & $-0.098 \;   (0.058)$          \\
log(street\_length\_avg) & $0.151 \;   (0.012)^{***}$     \\
log(nodes)               & $0.061 \;   (0.006)^{***}$     \\
IsPyramid $\Delta$       & $-0.223 \;   (0.030)^{***}$    \\
ICoast                   & $0.062 \;   (0.013)^{***}$     \\
IBorder                  & $-0.004 \;   (0.016)$          \\
\hline
R$^2$                    & $0.529$                        \\
Adj. R$^2$               & $0.528$                        \\
Num. obs.                & $5625$                         \\
\hline
\multicolumn{2}{l}{\scriptsize{$^{***}p<0.001$; $^{**}p<0.01$; $^{*}p<0.05$}}
\end{tabular}
\caption{Model polycentrism $\phi$ depending of the population, region, the mean building area, differences in elevation, the area covered by buildings with more than 100 m2 (a100), the footprint within the centre and the area of buildings, the average street length and the ratio between the average area of buildings within 1 km of the city centre and the average area of buildings in the city (IsPyramid $\Delta$), whether the city is coastal and whether the city is within an international border (less than 50 km).}
\label{TablePolcentrism}
\end{center} }
\end{table}

\subsection{Polycentrism in cities} 

Cities frequently develop and grow in a polycentric manner, that is, by merging different urban areas or adding new centres. Here, we measure polycentrism using the spatial distribution of buildings, the corresponding kernel density surface and the contour tree it forms. Results show that the number of branches of the corresponding contour tree increase with city size (Figure \ref{ApFigure}-B, C). However, those new branches add a small relative volume for small cities. In general, only cities above 100 thousand inhabitants are polycentric, where those extra branches add volume separate to the city's centre.

There is a wide variation in the levels of polycentrism in cities. Take, for example, an urban area with roughly one million inhabitants. The levels of polycentrism vary from 1 to values above 20, suggesting that it is not only size that affects the polycentric growth of cities. 

Sprawling cities are more polycentric. Even though greater sprawl signals larger distances between buildings, this does not necessarily translate into longer distances travelled by a city dweller in a polycentric city. Polycentric urban forms mean that multiple city centres exist so that people can meet their needs closer to home (e.g.,  the ``15-minute city'') rather than travelling to one centre (as in a monocentric city).

\subsection{Impact of city size on urban form} 

People from larger cities tend to produce more patents, have a higher income per capita, and suffer more crime and some types of diseases \cite{GrowthBettencourt}. People from larger cities migrate less and are more likely to return after moving \cite{ScalingMigrationRPC}. Smaller cities usually require more infrastructure per person, such as the road surface or the petrol stations \cite{GrowthBettencourt}. A mathematical expression to capture the impact of city size is obtained by adjusting the equation
\begin{equation}
Y_i = \alpha P_i^\beta,
\end{equation}
where $Y_i$ is the variable of interest for city $i$. Then, $Y_i$ could be the number of restaurants in the city $i$, for example. We obtain the coefficients $\alpha$ and $\beta$ through a regression. With $\hat{\beta}>1$ the results are called ``superlinear'' and indicate that large cities have higher values of the variable $Y$ per capita (since the per capita rate $Y_i/P_i$ is given by $\hat{\alpha} P_i^{\hat{\beta}-1}$). With $\hat{\beta}<1$ results are ``sublinear'' and with $\hat{\beta} \approx 1$ city size has little or no impact on the per capita rate of that city. Therefore, the coefficient of interest is usually $\beta$ and values above or below $\beta = 1$ are critical. It has been observed that for some social indicators, $\beta = 1.15$ is frequently obtained, which means that when comparing two cities, $i$ and $j$, where the population of $j$ is ten times the population of $i$, then, the expected values of $Y_j$ are approximately $10^{1.15} = 14.13$ times larger, and, on a per capita basis, city $j$ has 1.413 more of $Y$ than city $i$. Thus, comparing a city with 100,000 inhabitants and a city ten times larger, with one million people, it has 14.13 times more income, so a person from a large city has 1.413 times the income of a person from a smaller city, for example. Also, values of $\hat{\beta} \in (0, 1)$ indicate that the corresponding variable $Y$ increases with city size but at a slower rate than population (so the per capita ratio decreases with size). Finally, the same mechanism can be used to detect if some city indicator decreases with a larger population when $\hat{\beta} < 0$.

We consider scaling with respect to city size, so we fit the equations
\begin{eqnarray*}
\text{Buildings: }& & B_i = \alpha_B P_i ^ {\beta_B} \\
\text{Area of buildings: }& & A_i = \alpha_A P_i ^ {\beta_A} \\
\text{Sprawl: }& & S_i = \alpha_S P_i ^ {\beta_S}\\
\text{Elongation: }& &  E_i = \alpha_E P_i ^ {\beta_E} \\
\text{Distance: }& & D_i = \alpha_D P_i ^ {\beta_D} \\
\end{eqnarray*}
where $\alpha_B$ and $\beta_B$ are the scaling parameters which indicate whether city size impacts the number of buildings. Values of $\beta_B > 1$ indicate that the number of buildings in a city grows faster than its population. In other words, there are more per capita buildings in larger cities. And similarly, for the other coefficients, values above one would suggest larger buildings, more urban sprawl or elongated shapes. We fit similar coefficients also for the fragmentation, the observed average distance between buildings, the polycentrism, footprint and pyramid.

\begin{table}
{\scriptsize
\begin{center}
\begin{tabular}{ccc|c}
\hline
Variable & Scaling & cut-off & observed\\
\hline
Buildings & $B_i = \alpha_B P_i^{\beta_B}$ & $\beta_B = 1$ & 0.980997\\
Area      & $A_i = \alpha_A P_i^{\beta_A}$ & $\beta_A = 0$ & 0.067196\\
Sprawl     & $S_i = \alpha_S P_i^{\beta_S}$ & $\beta_S = 0$ & 0.010522\\
Elongation & $E_i = \alpha_E P_i^{\beta_E}$ & $\beta_E = 0$& 0.005021 \\
\hline
Fragmentation & $\psi_i = \alpha_\psi P_i^{\beta_\psi}$ & $\beta_\psi = 0$& 0.007772 \\
\hline
Distance       & $D_i = \alpha_D P_i^{\beta_D}$ & $\beta_D = 1/2$ & 0.531868\\
Distance$^\star$ & $D_i^\star = \alpha_{D^\star} P_i^{\beta_{D^\star}}$ & $\beta_{D^\star} = 1/2$ & 0.52410\\
\hline
Polycentrism   & $\phi_i = \alpha_\phi P_i^{\beta_\phi}$ & $\beta_\phi=0$ & 0.248818 \\
\hline
Footprint      & $\theta_i = \alpha_{\theta} P_i^{\beta_{\theta}}$ & $\beta_{\theta} = 1$ & 1.048 \\
Volume         & $V_i = \alpha_V P_i^{\beta_{V}}$ & $\beta_{V} = 1$ & 1.0238 \\
\hline
Pyramid       & $\Delta_i = \alpha_{\Delta} P_i^{\beta_{\Delta}}$ & $\beta_{\Delta} = 0$ & 0.024897 
\end{tabular}
\end{center}
}
\end{table}

Imagine a city with $n$ buildings, each occupied by one person, with the same area (say, 1 m2). If buildings are arranged one next to the other to minimise the average distance between them, they would be arranged roughly circular and compactly. For a sufficiently large number of buildings, they will form approximately a circle with an area of $n$. The mean distance between those buildings would be $128 /(45 \pi) n^{1/2}$. Therefore, in a city with a circular and compact shape, distances should grow with the square root of the population. In other words, if we fit the scaling coefficient $D_i = \alpha P_i^{\beta}$, a lower boundary of $\beta$ is given by $\beta \geq 1/2$. The BASE model is based on setting $D_i = \frac{128}{45 \pi} \sqrt{ B_i A_i S_i E_i}$, where we obtained the coefficient $\beta_D = 0.543$, so distances in cities grow at a faster rate than the boundary ($1/2$). The city metrics vary substantially across the five African regions. 

\subsection{The size of buildings across Africa} 

Most buildings across Africa have a tiny footprint. A small footprint corresponds to a low-rise construction. We only have some uncertainty on the height in buildings with a large footprint. Therefore, the distribution of the footprint of buildings is also an indicator of infrastructure, including construction materials, overcrowding, and resilient structures, among many economic indicators. Most buildings, particularly in small cities, are quite small (Table \ref{TableBuildingSizeH}). For example, most buildings in Bangui are smaller than 25 m2 (something like a square with 5 m on each side). We compare the distribution of building size between cities (Table \ref{TableLargeBuildings}). 

\begin{table}
{\tiny
\begin{center}
\begin{tabular}{l c}
\hline
 & $y = \log H$ \\
\hline
(Intercept)           & $1.6623 \; (0.0284)^{***}$  \\
log(Population\_2015) & $0.0204 \; (0.0021)^{***}$  \\
RegionEast            & $-0.0986 \; (0.0080)^{***}$ \\
RegionNorth           & $0.0917 \; (0.0080)^{***}$  \\
RegionSouth           & $-0.1279 \; (0.0098)^{***}$ \\
RegionWest            & $-0.0683 \; (0.0077)^{***}$ \\
log(Sprawl)           & $-0.0227 \; (0.0064)^{***}$ \\
log(Elongation)       & $-0.0186 \; (0.0079)^{*}$   \\
dif\_elev\_m          & $-0.0001 \; (0.0000)^{***}$ \\
ICoast                & $0.0505 \; (0.0064)^{***}$  \\
IBorder               & $-0.0274 \; (0.0082)^{***}$ \\
\hline
R$^2$                 & $0.3032$                    \\
Adj. R$^2$            & $0.3020$                    \\
Num. obs.             & $5583$                      \\
\hline
\multicolumn{2}{l}{\scriptsize{$^{***}p<0.001$; $^{**}p<0.01$; $^{*}p<0.05$}}
\end{tabular}
\caption{Average building size $H_i$ depending of the population, region, sprawl, elongation, differences in elevation, whether the city is coastal and whether the city is within an international border (less than 50 km).}
\label{TableBuildingSizeH}
\end{center}}
\end{table}

\begin{table}
{\tiny
\begin{center}
\begin{tabular}{l c}
\hline
 & $y = a100$ \\
\hline
(Intercept)              & $-0.1562 \;   (0.0209)^{***}$   \\
log(Population\_2015)    & $-0.0027 \;   (0.0020)$         \\
RegionEast               & $0.0472 \;   (0.0036)^{***}$    \\
RegionNorth              & $0.1099 \;   (0.0041)^{***}$    \\
RegionSouth              & $0.0935 \;   (0.0045)^{***}$    \\
RegionWest               & $0.0780 \;   (0.0037)^{***}$    \\
MeanBuildingArea         & $7678.7422 \; (201.9923)^{***}$ \\
log(dif\_elev\_m)        & $-0.0036 \;   (0.0012)^{**}$    \\
log(Polycentrism)        & $-0.0052 \;   (0.0031)$         \\
TotalFootprintCentre1km  & $0.0000 \;   (0.0000)$          \\
MeanArea1000m            & $0.0007 \;   (0.0002)^{***}$    \\
log(nodes)               & $0.0131 \;   (0.0013)^{***}$    \\
log(street\_length\_avg) & $0.0071 \;   (0.0028)^{*}$      \\
IsPyramid $\Delta$       & $0.0010 \;   (0.0070)$          \\
ICoast                   & $0.0130 \;   (0.0029)^{***}$    \\
IBorder                  & $0.0091 \;   (0.0037)^{*}$      \\
\hline
R$^2$                    & $0.8979$                        \\
Adj. R$^2$               & $0.8976$                        \\
Num. obs.                & $5625$                          \\
\hline
\multicolumn{2}{l}{\scriptsize{$^{***}p<0.001$; $^{**}p<0.01$; $^{*}p<0.05$}}
\end{tabular}
\caption{Area from buildings larger than 100 m depending of the population, region, the mean building area, differences in elevation, the footprint within the centre and the area of buildings, the average street length, the ratio between the average area of buildings within 1 km of the city centre and the average area of buildings in the city (IsPyramid $\Delta$), whether the city is coastal and whether the city is within an international border (less than 50 km).}
\label{TableLargeBuildings}
\end{center} }
\end{table}

However, outliers in terms of building size are relevant, and they do alter the city indicators. For example, consider two cities with 25,000 buildings (roughly 100,000 inhabitants) each, with a similar abundance of small buildings. Observing that one of those two cities has 250 large buildings (1\% of them) changes the way we look at that city since it could correspond to a state capital, for instance. Also, results vary if by ``large'' buildings we mean with a footprint of more than 100 m2 or more than 500 m2. Therefore, we consider the percentage of the built-up area constructed from buildings larger than some threshold $\kappa>0$. We vary the values of $\kappa$ from 20 m2 (tiny buildings) to 5000 m2. To detect the impact of city size on the number of large and small buildings, we consider a threshold area $\rho$ and measure, for city $i$, the number of buildings smaller than $\rho$, say $s_i(\rho)$, and the number of buildings larger than $\rho$, say $l_i(\rho)$. Then, we fit the Poisson regression
\begin{equation}
s_i(\rho) = \alpha_{s}(\rho) P_i^{\beta_s(\rho)}, \text{ and } l_i(\rho) = \alpha_{l}(\rho) P_i^{\beta_l(\rho)}.
\end{equation}
For example, values of $\beta_s(\rho)$ below 1 indicate an abundance of buildings smaller than $\rho$ in smaller cities, whereas values above 1 indicate an abundance of small buildings in larger cities. And the same of $\beta_l(\rho)$ for buildings larger than $\rho$. Different values of $\rho$ result in a different division of small or large buildings. Results show that the number of buildings in a city is sublinear (so a city with 100,000 inhabitants usually has more buildings per person than a city with one million people, ten times larger), but the relationship changes with building size (Figure \ref{ApFigure}-D, E). Considering buildings with more than 100 m2, we obtain that the scaling coefficient is approximately one, meaning the same number of buildings per person in small and large cities. As buildings grow, they become scarce in small cities, so the scaling coefficient increases. For buildings larger than 500 m2, the scaling coefficient is $\beta_{n500} = 1.12$, meaning that for a city that is ten times the size, the number of buildings larger than 500 m2 is $10^{1.12}=13.2$ times larger. Thus, larger cities have bigger buildings.

The footprint of a city constructed from buildings larger than a specific size is superlinear, and that relationship increases with the threshold size (Figure \ref{ApFigure}-E). The area of a city constructed of buildings larger than 500 m2, for example, shows a scaling coefficient of $\beta_{a500} = 1.15$, so again, if a city is ten times larger, then it has $10^{1.15} = 14.1$ times more surface of large buildings. 

\subsection{The uneven distribution of large buildings across a city} 

The presence of large buildings is related to city size, but the location of those large buildings within a city also plays a role. We distinguish some cities where the ``centre'' has a high frequency of large buildings, and the area decays as the distance to the city centre increases, thus, forming a ``pyramid'' type of city. In contrast, a different kind of city is observed when large buildings are spread across the urban polygon, likely signalling factories, logistics and storage rooms. The city centre is identified as the point with the highest weighted density of buildings, obtained by a kernel density estimate of the point process formed by the coordinates of buildings, weighted by their footprint. Thus, the density surface gives the constructed area within some radius at each $x,y$ coordinate. We then identify the city centre as the location with the highest built density. Although the city centre could be defined differently (for example, by analysing the polygon's centroid, looking for some central business district or looking at the building closer to others), our technique identifies the location in the city with the highest level of constructed surface and it is dependent only on buildings and not on the shape of the polygon. 

We construct a metric that captures whether a city has a pyramid or a more flat dispersion of buildings. But we also want to consider the relative size of buildings. That is, a city might be regarded as more pyramid-like if the largest buildings in the city are by its centre, even if those buildings are relatively small. For measuring a relative factor of how the city has a pyramidal distribution, we construct an index based on the relative size of buildings in cities. The coefficient is constructed by the ratio between the average size of buildings within the city centre and the average size of buildings in that city. Formally, let $\mu_i$ be the average size of buildings in city $i$, and let $\nu_i$ be the average size of buildings within a radius of 1 km within the city centre. Then, the flatness coefficient $\Delta_i$ is given by 
\begin{equation}
\Delta_i = \frac{\nu_i}{\mu_i},
\end{equation}
where $\Delta_i \approx 1$ means that buildings in a city have a similar footprint in the city centre. Larger values of $\Delta_i$ suggest a pyramid-type of the city, and smaller values of $\Delta_i$ suggest that the city centre has smaller buildings than the outskirts.

By comparing the average size of buildings within the densest point to the rest of the city, we obtain an index that identifies the dispersion of large buildings in a town. If the recognised centre has smaller buildings, it corresponds to a city centre with small but very compact buildings. If buildings within the city centre are of equal size to the rest of the city, it means a slight variability in the size of buildings, corresponding to a flat city. We consider the average size of buildings within a radius of 1 km within the city centre. The flatness coefficient $\Delta_i$, combined with other aspects such as city size, mean building size and emptiness, distinguishes the internal structure of cities (Table \ref{PyramidCities}).

\begin{table}
{\tiny
\begin{center}
\begin{tabular}{l c}
\hline
 & $y = \log \Delta$ \\
\hline
(Intercept)             & $-0.6059 \;   (0.0392)^{***}$    \\
log(Population\_2015)   & $0.0763 \;   (0.0039)^{***}$     \\
RegionEast              & $0.0364 \;   (0.0097)^{***}$     \\
RegionNorth             & $-0.0301 \;   (0.0097)^{**}$     \\
RegionSouth             & $0.0649 \;   (0.0128)^{***}$     \\
RegionWest              & $-0.0816 \;   (0.0097)^{***}$    \\
MeanBuildingArea        & $-1780.1001 \; (131.6482)^{***}$ \\
TotalFootprintCentre1km & $-0.0000 \;   (0.0000)^{***}$    \\
FootprintPerPerson      & $0.0032 \;   (0.0003)^{***}$     \\
\hline
R$^2$                   & $0.2331$                         \\
Adj. R$^2$              & $0.2320$                         \\
Num. obs.               & $5625$                           \\
\hline
\multicolumn{2}{l}{\scriptsize{$^{***}p<0.001$; $^{**}p<0.01$; $^{*}p<0.05$}}
\end{tabular}
\caption{Ratio building size inside the city centre, $\Delta_i$ depending on the population, region, the mean building area, the footprint within the centre and the footprint per person.}
\label{PyramidCities}
\end{center} }
\end{table}

We vary the ratio within the distance of the densest point of the city. Within a reasonable distance between 500 m and 3 km, we obtain similar results (Table \ref{Table1And3km} shows the results when considering a radius of 1 and 3 km).

\begin{table}
{\tiny
\begin{center}
\begin{tabular}{l c c}
\hline
 & $y = \log \theta_{1km}$ & $y = \log \theta_{3km}$ \\
\hline
(Intercept)           & $6.4191^{***}$     & $3.5406^{***}$     \\
                      & $(0.0864)$         & $(0.0859)$         \\
log(Population\_2015) & $0.5560^{***}$     & $0.8415^{***}$     \\
                      & $(0.0074)$         & $(0.0074)$         \\ 
RegionEast            & $0.2033^{***}$     & $0.1593^{***}$     \\
                      & $(0.0229)$         & $(0.0228)$         \\
RegionNorth           & $0.2968^{***}$     & $0.1778^{***}$     \\
                      & $(0.0259)$         & $(0.0258)$         \\
RegionSouth           & $0.2934^{***}$     & $0.5418^{***}$     \\
                      & $(0.0289)$         & $(0.0287)$         \\
RegionWest            & $0.2516^{***}$     & $0.1469^{***}$     \\
                      & $(0.0241)$         & $(0.0239)$         \\                      
MeanBuildingArea      & $-7007.4782^{***}$ & $-5026.2724^{***}$ \\
                      & $(1415.9295)$      & $(1406.6931)$      \\
log(Polycentrism)     & $-0.6074^{***}$    & $-0.6226^{***}$    \\
                      & $(0.0178)$         & $(0.0177)$         \\
log(dif\_elev\_m)     & $0.0486^{***}$     & $0.1111^{***}$     \\
                      & $(0.0071)$         & $(0.0070)$         \\

a100                  & $0.6553^{***}$     & $1.2729^{***}$     \\
                      & $(0.0826)$         & $(0.0821)$         \\
MeanArea1000m         & $0.0068^{***}$     & $0.0013$           \\
                      & $(0.0012)$         & $(0.0012)$         \\
Capital               & $-0.0937$          & $-0.3469^{***}$    \\
                      & $(0.0635)$         & $(0.0631)$         \\
IsPyramid $\Delta$    & $-0.3640^{***}$    & $-0.1067^{*}$      \\
                      & $(0.0432)$         & $(0.0429)$         \\
ICoast                & $0.0135$           & $0.0729^{***}$     \\
                      & $(0.0184)$         & $(0.0183)$         \\
\hline
R$^2$                 & $0.6233$           & $0.8130$           \\
Adj. R$^2$            & $0.6224$           & $0.8126$           \\
Num. obs.             & $5509$             & $5509$             \\
\hline
\multicolumn{3}{l}{\scriptsize{$^{***}p<0.001$; $^{**}p<0.01$; $^{*}p<0.05$}}
\end{tabular}
\caption{Ratio building size inside city centre comparing a distance of $R = 1$ and $R = 3$ km. The dependant variable depends on population, region, men building area, polycentrism, differences in elevation, the area covered by buildings with more than 100 m2 (a100), the area of buildings within the centre, whether the city is a country capital, the ratio between building size nearby the centre and in the rest of the urban polygon and whether the city is coastal. }
\label{Table1And3km}
\end{center} }
\end{table}

Observing an urban area only at its densest point gives a powerful indication of what the rest of the city looks like and what is its urban form. The sprawl of a city inside a circle, referred to as the emptiness $T_i^{(R)} \geq 1$, is defined as the ratio between the surface of the city centre and the constructed part. A city with a high emptiness $T_i^{(R)}$ will be highly sprawled within the rest of the city (Table \ref{AppTableCentreOnly}), but it will also be more elongated and, in turn, more fragmented. If the emptiness of a city is double, then the overall sprawl increases 55\%, the elongation increases 35\%, and therefore, the average distance between buildings across the whole city increases 45\%.

\begin{table}
{\tiny
\begin{center}
\begin{tabular}{l c c c}
\hline
 & $y = \log S$ & $y = \log E$ & $y = \log \psi$ \\
\hline
(Intercept)           & $-4.831^{***}$ & $-1.185^{***}$ & $-3.008^{***}$ \\
                      & $(0.086)$      & $(0.077)$      & $(0.073)$      \\
log(Population\_2015) & $0.282^{***}$  & $0.188^{***}$  & $0.235^{***}$  \\
                      & $(0.006)$      & $(0.005)$      & $(0.005)$      \\
$\log (T^{(R)}$         & $0.633^{***}$  & $0.436^{***}$  & $0.535^{***}$  \\
                      & $(0.010)$      & $(0.009)$      & $(0.008)$      \\                      
RegionEast            & $0.156^{***}$  & $0.163^{***}$  & $0.159^{***}$  \\
                      & $(0.019)$      & $(0.017)$      & $(0.016)$      \\
RegionNorth           & $0.057^{**}$   & $-0.163^{***}$ & $-0.053^{***}$ \\
                      & $(0.018)$      & $(0.017)$      & $(0.016)$      \\
RegionSouth           & $0.343^{***}$  & $-0.034$       & $0.155^{***}$  \\
                      & $(0.022)$      & $(0.020)$      & $(0.019)$      \\
RegionWest            & $0.005$        & $-0.110^{***}$ & $-0.052^{***}$ \\
                      & $(0.018)$      & $(0.016)$      & $(0.015)$      \\
\hline
R$^2$                 & $0.482$        & $0.430$        & $0.503$        \\
Adj. R$^2$            & $0.482$        & $0.430$        & $0.503$        \\
Num. obs.             & $5625$         & $5625$         & $5625$         \\
\hline
\multicolumn{4}{l}{\scriptsize{$^{***}p<0.001$; $^{**}p<0.01$; $^{*}p<0.05$}}
\end{tabular}
\caption{Impact of the emptiness of the densest point of a city on its sprawl, elongation and fragmentation. }
\label{AppTableCentreOnly}
\end{center} }
\end{table}

\subsection{Vertical growth and densification of cities} 

We analyse the vertical growth in cities in two manners. First on the aggregate surface of the city and second within 1 km of its densest point. We decompose the constructed volume of a city, $V_i$ as the product of its footprint $\theta_i$ and its average height $H_i$, so $V_i = \theta_i H_i$. We write the volume, footprint and average height as a function of the population of the city, and obtain $\alpha_V P_i^{\beta_V} \propto P_i ^{ \beta_{\theta}} P_i^{\beta_H}$, so that $\beta_V = \beta_{\theta} + \beta_H$. We obtain that $\beta_{\theta} = 1.0502 \pm 0.0065$ and $\beta_H = 0.0181 \pm 0.0022$. Therefore, larger cities have more surface per person and a slightly higher building height. Combined, both effects give $\beta_V = 1.06836 \pm 0.0087$, meaning that larger cities have a slightly higher volume per person than smaller cities, but this is mostly since the footprint is larger.

Looking only at the urban footprint at a distance $R=1$ km from its densest point, we can also decompose the volume as $V_i^{(R)} = \theta_i^{(R)} H_i^{(R)}$. We express the footprint, height and volume in terms of the population and obtain the corresponding coefficients, with $\beta_{V^{(R)}} = \beta_{\theta^{(R)}} + \beta_{H^{(R)}}$. We obtain that $\beta_{\theta^{(R)}} = 0.43766 \pm 0.0063$ and that $\beta_{H^{(R)}} = 0.0955 \pm 0.0155$. Therefore, a larger city has more footprint within the densest point and becomes more vertical. A city with ten times more population has buildings that are $10^{0.0955} = 1.25$, so 25\% taller nearby the densest point of the city. Combining both effects of taller buildings and more constructed surfaces within the densest point, we obtain that $\beta_{V^{(R)}} = 0.5332$, meaning within the densest point, the constructed infrastructure increases considerably in large cities. A city with ten times more population has $10^{0.5332} = 3.413$ times more volume within the centre. However, most of the additional infrastructure is due to cities' horizontal growth, not taller buildings. Thus, even cities with a high density can still gain from vertical growth \cite{angel2021anatomy}.

It is also possible to observe the vertical growth of cities by its highest point $I_i$ within its urban footprint. By writing the scaling equation $I_i = \alpha_I P_i^{\beta_I}$, we get that $\beta_I = 0.2375 \pm 0.0046$, suggesting that within a large city, there are usually taller buildings. Here we obtain that the tallest building of a city is 73\% taller within a city with ten times more population.

Finally, the availability of taller buildings and infrastructure are affected by specific urban indicators (Table \ref{TableVolumeCentreV}). More elongated and fragmented cities and cities near an international border tend to have shorter buildings and fewer infrastructure volumes. 

\begin{table}
{\tiny
\begin{center}
\begin{tabular}{l c}
\hline
 & $y = \log V^{(R)}$ \\
\hline
(Intercept)           & $8.0045 \; (0.1206)^{***}$  \\
log(Population\_2015) & $0.5521 \; (0.0088)^{***}$  \\
RegionEast            & $-0.1181 \; (0.0338)^{***}$ \\
RegionNorth           & $-0.4718 \; (0.0341)^{***}$ \\
RegionSouth           & $0.0651 \; (0.0416)$        \\
RegionWest            & $-0.1366 \; (0.0328)^{***}$ \\
log(Sprawl)           & $-0.5051 \; (0.0270)^{***}$ \\
log(Elongation)       & $-0.1494 \; (0.0336)^{***}$ \\
dif\_elev\_m          & $-0.0004 \; (0.0001)^{***}$ \\
ICoast                & $-0.0326 \; (0.0271)$       \\
IBorder               & $0.1279 \; (0.0349)^{***}$  \\
\hline
R$^2$                 & $0.5163$                    \\
Adj. R$^2$            & $0.5154$                    \\
Num. obs.             & $5583$                      \\
\hline
\multicolumn{2}{l}{\scriptsize{$^{***}p<0.001$; $^{**}p<0.01$; $^{*}p<0.05$}}
\end{tabular}
\caption{Volume within densest location $V^{(R)}$ depending of the population, region, sprawl, elongation, differences in elevation, whether the city is coastal or nearby an international border.}
\label{TableVolumeCentreV}
\end{center}}
\end{table}

\subsection{Distance and energy consumption} 

Infrastructure and urban form are strongly linked and alter the requirements for transport energy \cite{seto2014human}. The mean distance between buildings in a city is our proxy for the ``energy'' that the city requires for transportation. Given the average distance between buildings on city $i$, expressed as a function of the population by $D_i(P_i) = \alpha_D P_i^{\beta_D}$ we estimate that the total energy consumed for transport in the city is proportional to $T_i =\alpha_T P_i^{\beta_T} = \alpha_T P_i^{\beta_D+1}$ for some $\alpha_T > 0$. This way, $P_i$ experiences a commuting distance in the city that is proportional to $P_i^{\beta_D}$. Having decomposed (as in \cite{gudipudi2019efficient} for the emissions of a city and as in \cite{angel2020anatomy} for its density) the mean distances in cities (Equation 1 in the manuscript) we get that the distances grow with the population of a city, with exponent $\beta_D = (\beta_B + \beta_A + \beta_S + \beta_E)/2$ and the energy consumed for transport grows with the population with exponent $\beta_T = (\beta_B + \beta_A + \beta_S + \beta_E)/2 + 1$. Although we cannot estimate values of $\alpha_T$ based on our data, we observe that at a continental level, $\beta_D = 0.532$. Therefore, we estimate that the total energy consumed for transport in African cities grows superlinearly, with a coefficient of $1.532$. Thus, a city with ten times the population requires 34 times more energy in transport because the city is larger, so with more and bigger buildings, distances grow.

The BASE model enables us to predict the number of buildings and the size of a city with some population $P_d$ and also estimates a city's sprawl and elongation index. Most countries in Africa will double their 2020 population, some even before 2050, and some cities might reach a population of 80 or even 100 million inhabitants \cite{100MillionCities}. Assuming that population is the only determinant for cities, we analyse the expected evolution of distances and energy as their population increases \cite{pumain2004scaling, depersin2018global}. Due to the continent's demographic expansion and urbanisation process, most cities will double their population within decades. And when a city doubles in size, the total energy requirements in the continent will increase three times. Our estimate varies by region. In North Africa, a city that doubles its population will require 2.97 times more energy, but a city in Central Africa will only require 2.63 times more energy since cities have less sprawl and become less elongated as they increase in size. The opposite happens in North Africa, where larger cities tend to have higher levels of fragmentation. 

Further, we construct three different scenarios for cities that double their population. For the current scenario, we use the obtained values at a continental level of the coefficients $\beta_B$, $\beta_A$, $\beta_S$, and $\beta_E$, and we also consider the best-case scenario, where the values considered are the smallest observed across the five regions, consisting of a sublinear growth in the number of buildings, as observed in North Africa (with $_N\beta_B = 0.926$), with roughly the same building size, as observed in South Africa (with $_S\beta_A = 0.018$), with less sprawled and elongated cities, as observed in Central Africa (with $_C\beta_S = -0.156$ and $_C\beta_E = -0.109$). And similarly, we construct the worst-case scenario, where larger cities have more and larger buildings, but they also become more sprawled and elongated. The reasoning is that at least one of the five regions follows the best and worst scenarios. Results show that the coefficient for the best-possible scenario gives $\beta^{b}_D = 0.3395$, with $\beta^{b}_T = 1.3395$, and the coefficient for the worst-possible scenario gives $\beta^{w}_D = 0.6815$, with $\beta^{w}_T = 1.6815$. Thus, in the best-case scenario, when population doubles, distances increase $2^{0.3395} = 1.265$ times, and the commuting energy increase by $2^{1.3395} = 2.53$ times. In the worst-case scenario, distances will increase by $2^{0.6815} = 1.606$ times and the commuting energy will increase by $2^{1.6815} = 3.206$ times.

\subsection{Regional level}

Observed results at a regional level in North, West, East, Central and South Africa (Table \ref{AppRegionalTables}).

\begin{table}
{\tiny
\begin{tabular}{l c c c c c}
\hline
 & $y = log B_N$ & $y = log B_W$ & $y = log B_E$ & $y = log B_C$ & $y = log B_S$ \\
\hline
(Intercept)           & $-0.370^{**}$ & $-0.715^{***}$ & $-1.388^{***}$ & $-1.038^{***}$ & $-0.472^{**}$ \\
                      & $(0.135)$     & $(0.116)$      & $(0.107)$      & $(0.181)$      & $(0.148)$     \\
log(Population\_2015) & $0.926^{***}$ & $0.957^{***}$  & $1.042^{***}$  & $0.990^{***}$  & $1.005^{***}$ \\
                      & $(0.013)$     & $(0.011)$      & $(0.010)$      & $(0.017)$      & $(0.014)$     \\
\hline
R$^2$                 & $0.763$       & $0.792$        & $0.886$        & $0.889$        & $0.916$       \\
\hline
 & $y = log A_N$ & $y = log A_W$ & $y = log A_E$ & $y = log A_C$ & $y = log A_S$ \\
\hline
(Intercept)           & $-10.300^{***}$ & $-11.264^{***}$ & $-10.413^{***}$ & $-10.992^{***}$ & $-10.151^{***}$ \\
                      & $(0.095)$       & $(0.102)$       & $(0.072)$       & $(0.131)$       & $(0.109)$       \\
log(Population\_2015) & $0.044^{***}$   & $0.144^{***}$   & $0.035^{***}$   & $0.076^{***}$   & $0.018$         \\
                      & $(0.009)$       & $(0.010)$       & $(0.007)$       & $(0.012)$       & $(0.010)$       \\
\hline
R$^2$                 & $0.015$         & $0.100$         & $0.019$         & $0.082$         & $0.007$         \\
\hline
 & $y = log S_N$ & $y = log S_W$ & $y = log S_E$ & $y = log S_C$ & $y = log S_S$ \\
\hline
(Intercept)           & $-1.366^{***}$ & $-0.073$      & $-0.351^{*}$ & $1.528^{***}$  & $0.843^{***}$  \\
                      & $(0.128)$      & $(0.095)$     & $(0.136)$    & $(0.147)$      & $(0.192)$      \\
log(Population\_2015) & $0.106^{***}$  & $-0.026^{**}$ & $0.026^{*}$  & $-0.156^{***}$ & $-0.078^{***}$ \\
                      & $(0.012)$      & $(0.009)$     & $(0.013)$    & $(0.014)$      & $(0.019)$      \\
\hline
R$^2$                 & $0.045$        & $0.004$       & $0.003$      & $0.231$        & $0.038$        \\
\hline
 & $y = log E_N$ & $y = log E_W$ & $y = log E_E$ & $y = log E_C$ & $y = log E_S$ \\
\hline
(Intercept)           & $0.964^{***}$ & $1.909^{***}$ & $2.094^{***}$ & $3.150^{***}$  & $2.514^{***}$  \\
                      & $(0.100)$     & $(0.074)$     & $(0.108)$     & $(0.153)$      & $(0.200)$      \\
log(Population\_2015) & $0.071^{***}$ & $-0.017^{*}$  & $-0.001$      & $-0.109^{***}$ & $-0.065^{***}$ \\
                      & $(0.010)$     & $(0.007)$     & $(0.010)$     & $(0.015)$      & $(0.019)$      \\
\hline
R$^2$                 & $0.033$       & $0.003$       & $0.000$       & $0.120$        & $0.025$        \\
\hline
 & $y = log \psi_N$ & $y = log \psi_W$ & $y = log \psi_E$ & $y = log \psi_C$ & $y = log \psi_S$ \\
\hline
(Intercept)           & $-0.201$      & $0.918^{***}$ & $0.872^{***}$ & $2.339^{***}$  & $1.679^{***}$  \\
                      & $(0.108)$     & $(0.078)$     & $(0.113)$     & $(0.132)$      & $(0.189)$      \\
log(Population\_2015) & $0.089^{***}$ & $-0.022^{**}$ & $0.013$       & $-0.133^{***}$ & $-0.072^{***}$ \\
                      & $(0.010)$     & $(0.008)$     & $(0.011)$     & $(0.013)$      & $(0.018)$      \\
\hline
R$^2$                 & $0.044$       & $0.004$       & $0.001$       & $0.213$        & $0.033$        \\
\hline
 & $y = log \phi_N$ & $y = log \phi_W$ & $y = log \phi_E$ & $y = log \phi_C$ & $y = log \phi_S$ \\
\hline
(Intercept)           & $-2.959^{***}$ & $-1.791^{***}$ & $-3.168^{***}$ & $-0.999^{***}$ & $-2.435^{***}$ \\
                      & $(0.090)$      & $(0.058)$      & $(0.102)$      & $(0.080)$      & $(0.142)$      \\
log(Population\_2015) & $0.310^{***}$  & $0.186^{***}$  & $0.325^{***}$  & $0.103^{***}$  & $0.259^{***}$  \\
                      & $(0.009)$      & $(0.006)$      & $(0.010)$      & $(0.008)$      & $(0.014)$      \\
\hline
R$^2$                 & $0.448$        & $0.363$        & $0.458$        & $0.311$        & $0.441$        \\
\hline
 & $y = log D_N$ & $y = log D_W$ & $y = log D_E$ & $y = log D_C$ & $y = log D_S$ \\
\hline
(Intercept)           & $-5.635^{***}$ & $-5.170^{***}$ & $-5.128^{***}$ & $-3.776^{***}$ & $-3.732^{***}$ \\
                      & $(0.128)$      & $(0.085)$      & $(0.111)$      & $(0.130)$      & $(0.226)$      \\
log(Population\_2015) & $0.574^{***}$  & $0.528^{***}$  & $0.551^{***}$  & $0.400^{***}$  & $0.440^{***}$  \\
                      & $(0.012)$      & $(0.008)$      & $(0.011)$      & $(0.012)$      & $(0.022)$      \\
\hline
R$^2$                 & $0.579$        & $0.681$        & $0.672$        & $0.717$        & $0.474$        \\
\hline
 & $y = log D^\star_N$ & $y = log D^\star_W$ & $y = log D^\star_E$ & $y = log D^\star_C$ & $y = log D^\star_S$ \\
\hline
(Intercept)           & $-5.434^{***}$ & $-6.088^{***}$ & $-6.000^{***}$ & $-6.114^{***}$ & $-5.411^{***}$ \\
                      & $(0.072)$      & $(0.048)$      & $(0.057)$      & $(0.097)$      & $(0.094)$      \\
log(Population\_2015) & $0.485^{***}$  & $0.550^{***}$  & $0.539^{***}$  & $0.533^{***}$  & $0.512^{***}$  \\
                      & $(0.007)$      & $(0.005)$      & $(0.006)$      & $(0.009)$      & $(0.009)$      \\
\hline
R$^2$                 & $0.760$        & $0.879$        & $0.880$        & $0.891$        & $0.877$        \\
\hline
 & $y = log A100_N$ & $y = log A100_W$ & $y = log A100_E$ & $y = log A100_C$ & $y = log A100_S$ \\
\hline
(Intercept)           & $-1.590^{***}$ & $-2.886^{***}$ & $-2.477^{***}$ & $-3.744^{***}$ & $-1.412^{***}$ \\
                      & $(0.148)$      & $(0.134)$      & $(0.138)$      & $(0.254)$      & $(0.178)$      \\
log(Population\_2015) & $0.078^{***}$  & $0.201^{***}$  & $0.122^{***}$  & $0.204^{***}$  & $0.048^{**}$   \\
                      & $(0.014)$      & $(0.013)$      & $(0.013)$      & $(0.024)$      & $(0.017)$      \\
\hline
R$^2$                 & $0.019$        & $0.111$        & $0.061$        & $0.148$        & $0.017$        \\
\hline
 & $y = log FP1km_N$ & $y = log FP1km_W$ & $y = log FP1km_E$ & $y = log FP1km_C$ & $y = log FP1km_S$ \\
\hline
(Intercept)           & $8.233^{***}$ & $7.619^{***}$ & $8.033^{***}$ & $6.767^{***}$ & $8.250^{***}$ \\
                      & $(0.138)$     & $(0.100)$     & $(0.130)$     & $(0.201)$     & $(0.223)$     \\
log(Population\_2015) & $0.408^{***}$ & $0.471^{***}$ & $0.413^{***}$ & $0.515^{***}$ & $0.402^{***}$ \\
                      & $(0.013)$     & $(0.010)$     & $(0.012)$     & $(0.019)$     & $(0.022)$     \\
\hline
R$^2$                 & $0.375$       & $0.551$       & $0.454$       & $0.638$       & $0.435$       \\
\hline
 & $y = log FP_N$ & $y = log FP_W$ & $y = log FP_E$ & $y = log FP_C$ & $y = log FP_S$ \\
\hline
(Intercept)           & $3.146^{***}$ & $1.837^{***}$ & $2.015^{***}$ & $1.786^{***}$ & $3.193^{***}$ \\
                      & $(0.143)$     & $(0.096)$     & $(0.114)$     & $(0.193)$     & $(0.187)$     \\
log(Population\_2015) & $0.970^{***}$ & $1.100^{***}$ & $1.078^{***}$ & $1.066^{***}$ & $1.023^{***}$ \\
                      & $(0.014)$     & $(0.009)$     & $(0.011)$     & $(0.018)$     & $(0.018)$     \\
\hline
R$^2$                 & $0.760$       & $0.879$       & $0.880$       & $0.891$       & $0.877$       \\
\hline
 & $y = log SOpenGreen_N$ & $y = log SOpenGreen_W$ & $y = log SOpenGreen_E$ & $y = log SOpenGreen_C$ & $y = log SOpenGreen_S$ \\
\hline
(Intercept)           & $-0.543$      & $0.010$       & $-0.015$      & $2.414^{***}$ & $1.894^{***}$ \\
                      & $(0.489)$     & $(0.238)$     & $(0.252)$     & $(0.301)$     & $(0.482)$     \\
log(Population\_2015) & $1.018^{***}$ & $1.142^{***}$ & $1.163^{***}$ & $0.926^{***}$ & $1.030^{***}$ \\
                      & $(0.047)$     & $(0.023)$     & $(0.024)$     & $(0.029)$     & $(0.047)$     \\
\hline
R$^2$                 & $0.230$       & $0.561$       & $0.638$       & $0.717$       & $0.521$       \\
Num. obs.             & $1547$        & $1898$        & $1312$        & $416$         & $452$         \\
\hline
\multicolumn{6}{l}{\scriptsize{$^{***}p<0.001$; $^{**}p<0.01$; $^{*}p<0.05$}}
\end{tabular}
}
\caption{Coefficients across regions in Africa}
\label{AppRegionalTables}
\end{table}

\subsection{Terms summary}

Elements considered in the model (Table \ref{TermsSumm}).

\begin{table}
{\tiny
\begin{center}
\begin{tabular}{lp{35mm}p{45mm}}
$P_i \geq 10,000$ & population in city $i$ & count of people\\
$B_i \geq 2,000$ & number of buildings & count of buildings\\
$A_i \geq 10$ & average size of buildings & area in m2\\
$S_i \geq 0.25$ & factor for how measuring the sprawl of a city & coefficient\\ 
$E_i \geq 1$ & factor for how elongated is a city & coefficient greater than 1 \\
$D_i \geq 300$ & average distance between buildings & distance in 2\\
$D^\star_i \geq 300$ & average distance if the city had perfect packing & ideal distance in 2. \\
$\psi_i \geq 1$ & Fragmentation & computed as $D_i / D_i^\star$. \\
$\phi_i \geq 1$ & Polycentrism & weighted by the volume of each branch in the contour density of the city \\
$\theta_i \geq 60,000$ & Footprint & given by $B_i A_i$ in m2\\
$\Delta_i \geq 0$ & Pyramid & ratio between the average size of buildings within 1 km of the densest point of a city and the average size in that city $A_i$. \\
$T_i^{(R)} \geq 1$ & Emptiness & sprawl inside a circle of radius $R$ centred at the most dense location of the city.
\end{tabular}
\end{center} }
\caption{Terms summary}
\label{TermsSumm}
\end{table}

\bibliographystyle{unsrt}

\end{document}